\documentclass[aps,prd,showpacs,superscriptaddress,preprintnumbers]{revtex4}
\usepackage{graphicx,float,wrapfig,subfigure}
\usepackage{amsfonts,amsmath,amssymb,amstext}
\usepackage{latexsym}
 \usepackage{bm}
\usepackage{color}
\usepackage[normalem]{ulem}

\newcommand{\be}{\begin{equation}}
\newcommand{\ee}{\end{equation}}
\newcommand{\ba}{\begin{eqnarray}}
\newcommand{\ea}{\end{eqnarray}}

\newcommand{\sign}{\,\mbox{sign}}

\definecolor{red}{rgb}{0.7,0,0}
\definecolor{green}{rgb}{0,0.5,0}

\begin{document}

\title{Polarization tensor of magnetized quark-gluon plasma at nonzero baryon density}
\date{October 11, 2021}

\author{Xinyang Wang}
\email{wangxy@ujs.edu.cn}
\affiliation{Department of Physics, Jiangsu University, Zhenjiang 212013 P.R. China}

\author{Igor Shovkovy}
\email{igor.shovkovy@asu.edu}
\affiliation{College of Integrative Sciences and Arts, Arizona State University, Mesa, Arizona 85212, USA}
\affiliation{Department of Physics, Arizona State University, Tempe, Arizona 85287, USA}

\begin{abstract}
We derive a general expression for the absorptive part of the one-loop photon polarization tensor in a strongly magnetized quark-gluon plasma at nonzero baryon chemical potential. To demonstrate the  application of the main result in the context of heavy-ion collisions, we study the effect of a nonzero baryon chemical potential on the photon emission rate. The rate and the ellipticity of photon emission are studied numerically as a function the transverse momentum (energy) for several values of temperature and chemical potential. When the chemical potential is small compared to the temperature, the rates of the quark and antiquark splitting processes (i.e., $q\rightarrow q +\gamma$ and $\bar{q}\rightarrow \bar{q} +\gamma$, respectively) are approximately the same. However, the quark splitting gradually becomes the dominant process with increasing the chemical potential. We also find that increasing the chemical potential leads to a growing total photon production rate but has only a small effect on the ellipticity of photon emission. The quark-antiquark annihilation ($q+\bar{q}\rightarrow \gamma$) also contributes to the photon production, but its contribution remains relatively small for a wide range of temperatures and chemical potentials investigated. 
\end{abstract}

\maketitle

\section{introduction}

Relativistic plasmas appear when a matter is subject to extremely high temperatures or densities. Sufficiently high temperatures existed, for example, in the Early Universe soon after the Big Bang \cite{Applegate:1985qt,Gorbunov-Rubakov}. One can also produce a hot quark-gluon plasma in heavy-ion collision experiments, often called the Little Bangs \cite{Yagi:2005yb}. High-density relativistic plasmas, on the other hand, appear naturally inside compact stars \cite{Freedman:1977gz,Lattimer:2000nx,Steiner:2012xt}. Even electron quasiparticles can form relativistic-like plasmas in some topological semimetals \cite{DiracWeylSemimetals}. 

Strong magnetic fields are ubiquitous in cosmology \cite{Vachaspati:1991nm,Brandenburg:1996fc,Grasso:2000wj,Giovannini:2003yn,Boyarsky:2011uy}, astrophysics \cite{Thompson:1993hn,Cardall:2000bs,Price:2006fi}, and heavy-ion collisions \cite{Skokov:2009qp,Voronyuk:2011jd,Deng:2012pc}. They can drastically modify the thermodynamic and transport properties of relativistic plasmas. They also affect the emission and absorption properties, change the spectra of collective modes, etc. One of the characteristics that capture the effects of the magnetic field is the photon polarization tensor. Its real part, for example, determines the spectra of electromagnetically active collective modes. The imaginary part, on the other hand, is used in the calculation of the (optical) conductivity and the photon emission (absorption) rate. 

There has been substantial progress in studies of the polarization effects in magnetized plasmas in recent years. In the regime of  nonzero temperature, the polarization tensor was calculated in the lowest Landau level approximation \cite{Bandyopadhyay:2016fyd} and the weak-field limit \cite{Das:2019nzv,Ghosh:2019kmf}. Some results beyond the Landau level approximation were obtained as well \cite{Sadooghi:2016jyf,Ghosh:2018xhh,Ayala:2019akk,Ghosh:2020xwp,Ayala:2020wzl}. Among other results, a closed-form analytical expression for the absorptive part of the polarization tensor was derived in Ref.~\cite{Wang:2021ebh} (see also Ref.~\cite{Wang:2020dsr}) by using the Landau level representation for the fermion Green's function. The generalization of such studies to the case of a nonzero chemical potential was still missing, however. It is the purpose of this work to fill the corresponding knowledge gap. 

By following the same approach as in Refs.~\cite{Wang:2020dsr,Wang:2021ebh}, we will start by deriving the general expression for the photon polarization tensor and then concentrate on its absorptive part. Note that the latter includes the imaginary part of the symmetric tensor structure and the real part of the antisymmetric tensor structure. While a specific choice of the relativistic plasma is not crucial in this study, we will assume a two-flavor quark-gluon plasma for concreteness. Then, by using the absorptive part of the polarization tensor, we will calculate the differential photon production rate, which is relevant for heavy-ion physics. As we speculate, the qualitative features of photon emission could provide a measure of the magnetic field strength in the deconfined matter produced by heavy-ion collisions. Note that, despite the high transparency in relativistic collisions, the corresponding state of matter is characterized by a nonzero chemical potential \cite{Bass:1998ca}. The latter is an unavoidable artifact of the initial baryon charge carried by the colliding ions. This study aims to quantify the effect of a nonzero chemical potential on the direct photon emission from a strongly magnetized quark-gluon plasma.

This paper is organized as follows. We outline the derivations of the photon polarization tensor in Sec.~\ref{sec:Polarization} and the photon emission rate in Sec.~\ref{sec:Photon-emission}. The corresponding results generalize the work of Refs.~\cite{Wang:2020dsr,Wang:2021ebh} to the case of a nonzero chemical potential. The numerical results for the photon emission are presented in Sec.~\ref{sec:Photon-emission-numerical}. The summary and conclusions are given in Sec.~\ref{sec:Summary}. Several appendices at the end of the paper contain useful technical details.

\section{Polarization function with finite chemical potential}
\label{sec:Polarization}

The study in this paper is a continuation of the work started in Refs.~\cite{Wang:2020dsr,Wang:2021ebh}. Here we extend the photon polarization tensor of a magnetized quark-gluon plasma to the case of a nonzero baryon chemical potential. The calculations are at the leading-order one-loop approximation. It is a good approximation at sufficiently strong magnetic fields and high temperatures when the subleading corrections of order $\alpha\alpha_s$ are under control. Note that $\alpha=1/137$ is the fine structure constant, while $\alpha_s$ is the QCD coupling defined at a relevant physics scale (e.g., temperature, chemical potential, and/or magnetic field). After adjusting the electric charges and masses of particles, the result will be also valid for the QED plasma. In such a case, the validity of the one-loop approximation will be excellent because the subleading corrections of order  $\alpha^2$ are negligible. 

For simplicity, we assume that the masses of both light quarks are the same, i.e., $m_f =m=5~\mbox{MeV}$, where $f=u,d$. We define $e_f = eq_f$ as the flavor-dependent quark charge, where $q_{u} = 2/3$, $q_d =-1/3$, and  $e$ is the absolute value of the electron charge. We choose the magnetic field $\mathbf{B}$ to point in the $+z$ direction. The corresponding vector potential is taken in the Landau gauge, i.e., $\mathbf{A}=(-B y,0,0)$.

In a mixed coordinate-momentum space representation, the translation invariant part of the quark propagator $\bar{G}_f$ is given by~\cite{Miransky:2015ava}:
\ba
\bar{G}_f(t;\mathbf{r})&=& \int \frac{d\omega dp_z }{(2\pi)^2} e^{-i\omega t +i p_z z}
\bar{G}_f(\omega; p_z ;\mathbf{r}_\perp) ,
\ea
where 
\begin{equation}
\bar{G}_f(\omega, p_z ;\mathbf{r}_\perp) = i\frac{e^{-\mathbf{r}_\perp^2/(4\ell_{f}^2)}}{2\pi \ell_{f}^2}
\sum_{n=0}^{\infty}
\frac{\tilde{D}^f_{n}(\omega,p_z ;\mathbf{r}_\perp)}{(\omega+\mu)^2-p_z^2-m^2-2n|e_fB|},
\label{GDn-alt}
\end{equation}
and $\mu$ is the baryon chemical potential. We used the following shorthand notation for the numerator of the $n$th Landau level contribution:
\begin{equation}
\tilde{D}_{n}^f(\omega,p_z ;\mathbf{r}_\perp) = \left[(\omega+\mu)\gamma^0 -p^{3}\gamma^3 + m \right]
\left[{\cal P}^f_{+}L_n\left(\frac{\mathbf{r}_{\perp}^2}{2\ell_{f}^{2}}\right)
+{\cal P}^f_{-}L_{n-1}\left(\frac{\mathbf{r}_{\perp}^2}{2\ell_{f}^{2}}\right)\right]
-\frac{i}{ \ell_{f}^2}(\mathbf{r}_{\perp}\cdot\bm{\gamma}_{\perp}) 
 L_{n-1}^1\left(\frac{\mathbf{r}_{\perp}^2}{2 \ell_{f}^{2}}\right),
\end{equation}
where $\mathbf{r}_{\perp} = (x,y)$ is the position vector in the transverse (with respect to the magnetic field) plane, $L_{n}^\alpha(z)$ is the generalized Laguerre polynomial, ${\cal P}^f_{\pm}\equiv \frac12 \left(1\pm i s^f_\perp \gamma^1\gamma^2\right)$  are spin projectors, and $\ell_{f}=\sqrt{1/|e_f B|}$ is the flavor-specific  magnetic length. By definition, $s_\perp^f=\sign (e_f B)$ and $L_{-1}^\alpha(z) \equiv 0$. 

The photon polarization tensor in momentum space reads~\cite{Wang:2021ebh}
\begin{equation}
\Pi^{\mu\nu}(i\Omega_m;\mathbf{k}) =  4\pi N_c \sum_{f = u, d} \alpha_f T \sum_{k=-\infty}^{\infty} \int \frac{dp_z}{2\pi} 
\int d^2 \mathbf{r}_\perp e^{-i \mathbf{r}_\perp\cdot \mathbf{k}_\perp} 
\mbox{tr} \left[ \gamma^\mu \bar{G}_{f}(i\omega_k, p_z ;\mathbf{r}_\perp)  
\gamma^\nu \bar{G}_{f}(i\omega_k-i\Omega_m, p_z-k_z; -\mathbf{r}_\perp)\right],
\label{Pi_Omega_k-alt}
\end{equation}
where $\alpha_f = q_f^2 \alpha$ and  $\alpha= e^2/(4\pi)$ is the fine structure constant, $N_c =$ 3 is the number of colors, and the trace runs over the Dirac indices.  By using the standard convention, the fermionic and bosonic Matsubara frequencies are given by $\omega_k = (2k +1)\pi T$ and $\Omega_m = 2 m \pi T$, respectively. 

By substituting the fermion propagator in the Landau-level representation (\ref{GDn-alt}) into Eq.~(\ref{Pi_Omega_k-alt}) and performing the Matsubara sum with the help of Eq.~(\ref{Matsubara-sum}), we derive the following expression for the polarization function: 
\begin{equation}
\Pi^{\mu\nu}(i\Omega_m;\mathbf{k}) =
- \sum_{f = u,d}\frac{\alpha_f N_c}{\pi \ell_{f}^4} \sum_{n,n^\prime=0}^{\infty}\int \frac{dp_z}{2\pi} 
\sum_{\lambda, \eta =\pm 1} 
\frac{ \left[ n_F(E_{n,p_z,f}+\eta \mu)-n_F(\lambda E_{n^{\prime},p_z-k_z,f}+\eta \mu ) \right]}{4 \lambda  E_{n,p_z,f}E_{n^{\prime},p_z-k_z,f}\left[(E_{n,p_z,f}-\lambda E_{n^{\prime},p_z-k_z,f})+i \eta\Omega_m\right]}
\sum_{i=1}^{4}I_{i,f}^{\mu\nu}, 
\label{Pi_deriv-1}
\end{equation}
where $E_{n,p_z,f}= \sqrt{m^2+p_z^2 + 2 n |e_f B|}$ are Landau-level energies and $I_{i,f}^{\mu\nu}$ are tensor functions defined in Eqs.~(\ref{I_if-1-appD})--(\ref{I_if-4-appD}). Note that the result has the same general structure as in the $\mu=0$ case \cite{Wang:2020dsr,Wang:2021ebh}. However, the fermion distribution functions depend on the chemical potential now. Since the energies of quarks and antiquarks are shifted by $\pm\mu$ inside the distribution functions, the charge conjugation symmetry is broken explicitly. 

After replacing $i \Omega_m \to \Omega+ i \epsilon$, it is straightforward to extract the absorptive part of the retarded polarization tensor. The result reads
\begin{eqnarray}
\mbox{Im} \left[\Pi_R^{\mu\nu}(\Omega+i \epsilon;\mathbf{k}) \right] &=&
\sum_{f =u, d}\frac{\alpha_f N_c}{\ell_{f}^4} \sum_{n,n^\prime=0}^{\infty}\int \frac{dp_z}{2\pi} 
\sum_{\lambda, \eta =\pm 1} 
\frac{ \left[ n_F(E_{n,p_z,f}+\eta \mu)-n_F(\lambda E_{n^{\prime},p_z-k_z,f}+\eta \mu ) \right]}{4 \lambda \eta  E_{n,p_z,f}E_{n^{\prime},p_z-k_z,f}}\nonumber\\
&\times&
\sum_{i=1}^{4}I_{i,f}^{\mu\nu}
\delta\left(E_{n,p_z,f}-\lambda E_{n^{\prime},p_z-k_z,f}+\eta \Omega\right) .
\label{Im-Pol-fun}
\end{eqnarray}
(Strictly speaking, the notation is not precise since the expression gives the absorptive part of the tensor that includes both  imaginary part of the symmetric tensor structures and real part of antisymmetric ones.) Finally, by making use of the $\delta$-function and performing the integration over $p_z$, we derive the expression for the absorptive part of the polarization tensor:
\begin{equation}
\mbox{Im} \left[\Pi_R^{\mu\nu}(\Omega;\mathbf{k}) \right] =
\sum_{f =u,d}  \frac{\alpha_f N_{c}}{4\pi \ell_{f}^4} \sum_{n,n^\prime=0}^{\infty} 
\sum_{\lambda,\eta=\pm 1}\sum_{s=\pm 1} \Theta_{\lambda, \eta}^{n,n^{\prime}}(\Omega,k_z)
\frac{n_F(E_{n,p_z,f}+\eta \mu)-n_F(\lambda E_{n^{\prime},p_z-k_z,f}+\eta \mu) }{\eta\lambda  \sqrt{ \left( \Omega^2-k_z^2-(k_{-}^f)^2 \right) \left( \Omega^2-k_z^2-(k_{+}^f)^2\right)} }\sum_{i=1}^{4}I_{i,f}^{\mu\nu} \Bigg|_{p_z = p_{z,f}^{(s)}} ,
\label{Im-Pol-fun}
\end{equation} 
where the threshold function $\Theta_{\lambda, \eta}^{n,n^{\prime}}(\Omega,k_z)$ is defined as follows:
\begin{equation}
\Theta_{\lambda, \eta}^{n,n^{\prime}}(\Omega,k_z) = \left\{
\begin{array}{lll}
  \theta\left((k_{-}^f)^2 + k_z^2 - \Omega^2 \right) & \mbox{for}& \lambda=1, ~ \eta=-1, ~ n>n^{\prime},\\
  \theta\left( (k_{-}^f)^2 + k_z^2 - \Omega^2 \right) & \mbox{for}& \lambda=1, ~ \eta=1, ~ n<n^{\prime},\\
  \theta\left( \Omega^2 - k_z^2 -(k_{+}^f)^2 \right) & \mbox{for}& \lambda=-1, ~ \eta=-1,
  \end{array}
\right.
\end{equation} 
and $\Theta_{\lambda, \eta}^{n,n^{\prime}}(\Omega,k_z) =0$ otherwise. By definition, $\theta(x)$ is the Heaviside step function and the momentum thresholds are 
\begin{equation}
\label{kpm}
k_{\pm}^{f} = \left|\sqrt{m^2+2n|e_fB|} \pm \sqrt{m^2+2n^{\prime}|e_fB|}\right| .
\end{equation}
The solutions for the longitudinal momenta $p_{z}$, satisfying the energy conservation equation $E_{n,p_z,f}-\lambda E_{n^{\prime},p_z-k_z,f}+ \eta\Omega=0$, are given by the following explicit expressions~\cite{Wang:2020dsr,Wang:2021ebh}:
\begin{equation}
p_{z,f}^{(\pm)} = \frac{k_z}{2}\left[1+ \frac{2(n-n^{\prime})|e_fB|}{\Omega^2-k_z^2} 
\pm \frac{\Omega}{|k_z|}  \sqrt{\left(1-\frac{(k_{-}^{f})^2}{\Omega^2-k_z^2} \right)
\left( 1-\frac{(k_{+}^{f})^2}{\Omega^2-k_z^2}\right)} \right].
\label{pz-solution}
\end{equation}
From the energy conservation equation $E_{n,p_z,f}+ \eta\Omega = \lambda E_{n^{\prime},p_z-k_z,f}$, we find that
\begin{equation}
\label{energy-derivation}
2 \eta \Omega E_{n,p_z,f}= E_{n^{\prime},p_z-k_z,f}^2 - E_{n,p_z,f}^2 - \Omega^2
=2(n^{\prime}-n)|e_fB| -2 p_zk_z +k_z^2- \Omega^2.
\end{equation}
This allows us to solve for one of the fermions energies, 
\begin{equation}
 \left. E_{n,p_z,f}\right|_{p_z = p_{z,f}^{(\pm)} } = - \frac{\eta \Omega}{2} \left[
1+\frac{2(n-n^{\prime})|e_fB|}{\Omega^2-k_z^2}\pm \frac{|k_z|}{\Omega}\sqrt{ \left(1-\frac{(k_{-}^{f})^2}{\Omega^2-k_z^2} \right)\left( 1-\frac{(k_{+}^{f})^2}{\Omega^2-k_z^2}\right)} 
\right], \label{E1-solution}
\end{equation}
where we used the explicit expression for $p_{z,f}^{(\pm)}$ given in Eq.~(\ref{pz-solution}). 
By making use of the energy conservation equation once again, we also find the other energy, \begin{equation}
 \left. E_{n^{\prime},p_z-k_z,f} \right|_{p_z = p_{z,f}^{(\pm)} } =\frac{\lambda \eta \Omega}{2} \left[
1-\frac{2(n-n^{\prime})|e_fB|}{\Omega^2-k_z^2}\mp \frac{|k_z|}{\Omega}\sqrt{ \left(1-\frac{(k_{-}^{f})^2}{\Omega^2-k_z^2} \right)\left( 1-\frac{(k_{+}^{f})^2}{\Omega^2-k_z^2}\right)} 
\right].
\label{E2-solution}
\end{equation}
After combining all contributions to the absorptive part of $\Pi_R^{\mu\nu}(\Omega;\mathbf{k})$ and simplifying the final expression, we find that the polarization tensor has the following structure:
\begin{eqnarray}
\Pi_R^{\mu\nu}(\Omega;\mathbf{k}) &=&
 \left( \frac{k_\parallel^\mu k_\parallel^\nu }{k_\parallel^2} -  g_{\parallel}^{\mu\nu} \right)  \Pi_{1}
+ \left(g_{\perp}^{\mu\nu}+ \frac{k_\perp^\mu k_\perp^\nu}{k_\perp^2} \right) \Pi_{2}
+ \left( \frac{k_\parallel^\mu \tilde{k}_\parallel^\nu +\tilde{k}_\parallel^\mu k_\parallel^\nu}{k_\parallel^2}  
+ \frac{\tilde{k}_\parallel^\mu k_\perp^\nu +k_\perp^\mu \tilde{k}_\parallel^\nu}{k_\perp^2}\right) \Pi_{3}
\nonumber\\
&+&\left(\frac{k_\parallel^\mu k_\perp^\nu +k_\perp^\mu k_\parallel^\nu}{k_\parallel^2 } + \frac{k_\perp^2}{k_\parallel^2 }  g_{\parallel}^{\mu\nu}  - g_{\perp}^{\mu\nu} \right) \Pi_{4} 
+ \left( \frac{F^{\mu\nu} }{B}
+ \frac{k_\parallel^\mu \tilde{k}_{\perp}^\nu -\tilde{k}_{\perp}^\mu k_\parallel^\nu}{k_\parallel^2} \right) \tilde{\Pi}_{5}
+ \frac{\tilde{k}_\parallel^\mu \tilde{k}_{\perp}^\nu -\tilde{k}_{\perp}^\mu \tilde{k}_\parallel^\nu}{k_\parallel^2} \tilde{\Pi}_{6} ,
\label{ImPi-tensor-txt1}
\end{eqnarray}
where we used the following notation:
\begin{equation}
\begin{array}{lll}
g_{\parallel}^{\mu\nu} = \mbox{diag}(1,0,0,-1), \quad & 
k_\parallel^\mu = g_{\parallel}^{\mu\nu}k_{\nu}=k_0\delta^\mu_{0}+k_z\delta^\mu_{3},  \quad & 
\tilde{k}_\parallel^\mu  = - \varepsilon^{\mu 1 2 \nu}k_{\nu}= k_z \delta^{\mu}_{0} +k_0 \delta^{\mu}_{3}, \\
g_{\perp}^{\mu\nu} = \mbox{diag}(0,-1,-1,0),  \quad &  
k_\perp^\mu = g_{\perp}^{\mu\nu}k_{\nu}= k_x\delta^\mu_{1}+k_y\delta^\mu_{2}, \quad & 
\tilde{k}_{\perp}^{\mu}=-\varepsilon^{0 \mu \nu 3}k_{\nu}= k_{y} \delta^{\mu}_{1} -k_x \delta^{\mu}_{2}.\\
\end{array}
\label{ggkkkk}
\end{equation}
Note that $\tilde{k}_{\perp,\mu} \tilde{k}_{\perp}^{\mu} = k_{\perp,\mu} k_{\perp}^{\mu} = -k_\perp^2$,  
$\tilde{k}_{\parallel,\mu} \tilde{k}_\parallel^\mu = -k_{\parallel,\mu} k_\parallel^\mu= - k_\parallel^2$, and
$k_\mu \tilde{k}_{\perp}^{\mu} =k_{\mu} \tilde{k}_\parallel^\mu =0$. (For the photon energy, we use the notations $k_0$ and $\Omega$ interchangeably.)

The polarization tensor (\ref{ImPi-tensor-txt1}) has the same four {\em symmetric} tensor structures and two {\em antisymmetric} ones as in the $\mu=0$ case~\cite{Wang:2021ebh}. However, all component functions depend on $\mu$ now. Also, the antisymmetric terms, defined by the component functions $\tilde{\Pi}_{5}$ and $\tilde{\Pi}_{6}$, do not vanish because the charge conjugation symmetry is broken when $\mu\neq 0$.  

\section{Photon emission rate and ellipticity}
\label{sec:Photon-emission}

In this section we use the absorptive part of the polarization tensor to derive the expression for the photon production rate in a magnetized quark-gluon plasma at a nonzero baryon chemical potential. By definition, the rate is given by \cite{Kapusta:2006pm}
\begin{equation}
k^0\frac{d^3R}{dk_x dk_y dk_z}=-\frac{1}{(2\pi)^3}\frac{\mbox{Im}\left[\Pi^{\mu}_{R,\mu}(k)\right]}{\exp\left(\frac{k_0}{T}\right)-1}.
\label{diff-rate-2}
\end{equation}
In addition to the rate itself, it is also interesting to study the ellipticity of the photon emission. 
The conventional measure of ellipticity is quantified by 
\begin{equation}
v_2(k_T) = - \frac{1}{(2\pi)^3\mathcal{R}}  \int_0^{2\pi}\frac{ \mbox{Im}\left[\Pi^{\mu}_{R,\mu}(k)\right]}{\exp\left(\frac{k_0}{T}\right)-1} \cos(2\phi)d\phi,
\label{v2}
\end{equation}
where $\phi$ is the angle between the photon momentum $\mathbf{k}$ and the reaction plane. The 
normalization factor $\mathcal{R}$ is defined by
\begin{equation}
\mathcal{R} = - \frac{1}{(2\pi)^3 } \int_0^{2\pi}\frac{ \mbox{Im}\left[\Pi^{\mu}_{R,\mu}(k)\right]}{\exp\left(\frac{k_0}{T}\right)-1} d\phi .
\label{Integrated-rate}
\end{equation}
By following the same approach as in Refs.~\cite{Wang:2020dsr,Wang:2021ebh}, it is straightforward to obtain the imaginary part of the Lorentz contracted polarization tensor $\mbox{Im}\left[\Pi^{\mu}_{R,\mu}(k)\right]$ from Eq.~(\ref{Im-Pol-fun}). The result reads
\begin{eqnarray}
\mbox{Im} \left[\Pi^{\mu}_{R,\mu}\right] &=&
 \sum_{f = u,d}\frac{N_c \alpha_f}{4\pi  \ell_{f}^4} \sum_{n>n^\prime}^{\infty} \Theta_1
\sum_{s=\pm 1} 
\frac{n_F(E_{n^{\prime},p_z-k_z,f}-\mu) -n_F(E_{n,p_z,f}-\mu)}{\sqrt{ [( k_{-} ^f )^2 - k_y^2 ][ (k_{+} ^f)^2- k_y^2] } } 
\left(\mathcal{F}_1^f+\mathcal{F}_4^f \right)
\Bigg|_{p_z = p_{z,f}^{(s)} , \lambda =1, \eta =-1}
\nonumber\\
&+& \sum_{f = u,d}\frac{N_c \alpha_f}{4\pi  \ell_{f}^4} \sum_{n<n^\prime}^{\infty} \Theta_1
\sum_{s=\pm 1} 
\frac{n_F(E_{n,p_z,f}+\mu)-n_F(E_{n^{\prime},p_z-k_z,f}+\mu) }{\sqrt{ [( k_{-}^f  )^2 - k_y^2 ][ (k_{+}^f )^2- k_y^2] } }\left(
\mathcal{F}_1^f+\mathcal{F}_4^f \right)\Bigg|_{p_z = p_{z,f}^{(s)}  , \lambda =\eta =1}
\nonumber\\
&+&\sum_{f = u,d}\frac{N_c \alpha_f}{4\pi \ell_{f}^4} \sum_{n,n^\prime=0}^{\infty} \Theta_2
\sum_{s=\pm 1} 
\frac{n_F(E_{n,p_z,f}-\mu) +n_F(E_{n^{\prime},p_z-k_z,f}+\mu) -1}{\sqrt{ [k_y^2-( k_{-}  ^f)^2 ][ k_y^2-(k_{+}^f )^2]} }\left(
\mathcal{F}_1^f+\mathcal{F}_4^f \right) \Bigg|_{p_z = p_{z,f}^{(s)} , \lambda =\eta =-1},
\label{Im-Pi-19}
\end{eqnarray}
where used the notations: $\Theta_1 =  \theta\left((k_{-}^f)^2+k_z^2-\Omega^2\right)$ and 
$\Theta_2 = \theta\left(\Omega^2-k_z^2-(k_{+}^f)^2\right) $. We also introduced the following Lorentz contracted functions: $\mathcal{F}_i^{f} = g_{\mu\nu}I_{i,f}^{\mu\nu}$. Their explicit expressions are given in Eqs.~(\ref{F1-orig}) -- (\ref{F4-orig}) in Appendix~\ref{AP-tr}. Out of the four functions, only $\mathcal{F}_1^f$ and $\mathcal{F}_4^f $ are nontrivial.

For numerical calculations, it is convenient to rewrite the final expression as follows:
\begin{eqnarray}
\mbox{Im} \left[\Pi^{\mu}_{R,\mu}\right] &=&\sum_{f=u,d}
  \frac{N_c\alpha_f }{4\pi  \ell_{f}^4} \sum_{n>n^\prime}^{\infty}  
\frac{
\left[
\left[g(n, n^{\prime})+g(n^{\prime},n)\right] \theta\left((k_{-}^f)^2+k_z^2-\Omega^2\right)
-2 g(n, n^{\prime}) \theta\left(\Omega^2-k_z^2-(k_{+}^f)^2\right) \right]
 }{\sqrt{ \left( (k_{-}^f)^2+k_z^2-\Omega^2 \right)\left(( k_{+}^f)^2+k_z^2-\Omega^2\right) } } \left(
\mathcal{F}_1^f+\mathcal{F}_4^f \right)
\nonumber\\
&-& \sum_{f=u,d} \frac{N_c\alpha_f }{4\pi  \ell_{f}^4} \sum_{n=0}^{\infty} 
\frac{g_0(n)\theta\left(\Omega^2-k_z^2-(k_{+}^f)^2\right)}{\sqrt{ \left(\Omega^2-k_z^2\right)\left( \Omega^2-k_z^2-(k_{+}^f)^2 \right)} }\left(
\mathcal{F}_1^f+\mathcal{F}_4 ^f\right)  ,
\label{imag}
\end{eqnarray}
where 
\begin{equation}
\mathcal{F}_1^f +\mathcal{F}_4^f = 8\pi \left(n+n^{\prime}+m^2\ell_{f}^2\right)\left[\mathcal{I}_{0,f}^{n,n^{\prime}}(\xi)+\mathcal{I}_{0,f}^{n-1,n^{\prime}-1}(\xi) \right] 
- 8\pi  \left(n+n^{\prime}- \frac{\Omega^2-|\mathbf{k}|^2}{2} \ell_{f}^2 \right)
\left[\mathcal{I}_{0,f}^{n,n^{\prime}-1}(\xi)+\mathcal{I}_{0,f}^{n-1,n^{\prime}}(\xi) \right],
\label{F4-orig}
\end{equation}
and
\begin{eqnarray}
g(n, n^{\prime}) &=& 2-\sum_{s_1,s_2=\pm}
n_F\left[\frac{\Omega}{2}  -s_1 \mu 
+s_1 \frac{\Omega(n-n^{\prime})|e_f B|}{\Omega^2-k_z^2}+s_2 \frac{|k_z|}{2}\sqrt{ \left(1-\frac{(k_{-}^{f})^2}{\Omega^2-k_z^2} \right)\left( 1-\frac{(k_{+}^{f})^2}{\Omega^2-k_z^2}\right) }\right],\\ 
g_0(n) &\equiv& g(n, n) = 2-\sum_{s_1,s_2=\pm}
n_F\left(\frac{\Omega}{2} - s_1 \mu +s_2 \frac{|k_z|}{2}\sqrt{ 1-\frac{4(m^2+2n |e_fB|)^2}{\Omega^2-k_z^2} }
\right) .
\end{eqnarray}
We use this result in the next section to analyze the differential photon emission rate numerically.

\section{Numerical results}
\label{sec:Photon-emission-numerical}

In this section, we calculate the photon emission rate and the photon ellipticity in a strongly magnetized quark-gluon plasma at $\mu\neq 0$ numerically by using Eq.~(\ref{diff-rate-2}) and Eq.~(\ref{v2}) with the imaginary part in Eq.~(\ref{imag}). We assume that $x$-$y$ is the reaction plane and the beam direction is along the $x$-axis. This is a self-consistent configuration for noncentral collisions at mid-rapidity, where the magnetic field direction is (approximately) perpendicular to the reaction plane. At mid-rapidity, we can also set $k_x =0$. The remaining two components of the photon momentum are parametrized as follows: $k_y =k_T \cos\phi $ and $k_z =k_T\sin\phi $, where $k_T$ is the transverse momentum (with respect to the beam) and $\phi$ is the angle measured from the reaction plane. Note that the photon transverse momentum $k_T$ is the same as its energy $\Omega$ when $k_x =0$. 

Here we will consider a magnetized quark-gluon plasma with the same representative choices of the magnetic field strength (i.e.,  $|eB| = m_{\pi}^2$ and $|eB|=5m_{\pi}^2$) and temperature  (i.e.,  $T = 0.2~\mbox{GeV}$ and  $T = 0.35~\mbox{GeV}$) as in Ref.~\cite{Wang:2020dsr}. By following the standard convention, we give the values of the field in units of $m_\pi^2$, where $m_{\pi} = 0.135~\mbox{GeV}$. In conventional units, the two values of the field correspond to $B \approx 3.08\times 10^{18}~\mbox{G}$ and $B \approx 1.54\times 10^{19}~\mbox{G}$, respectively.

\begin{figure}[t]
\centering
\subfigure[]{\includegraphics[width=0.45\textwidth]{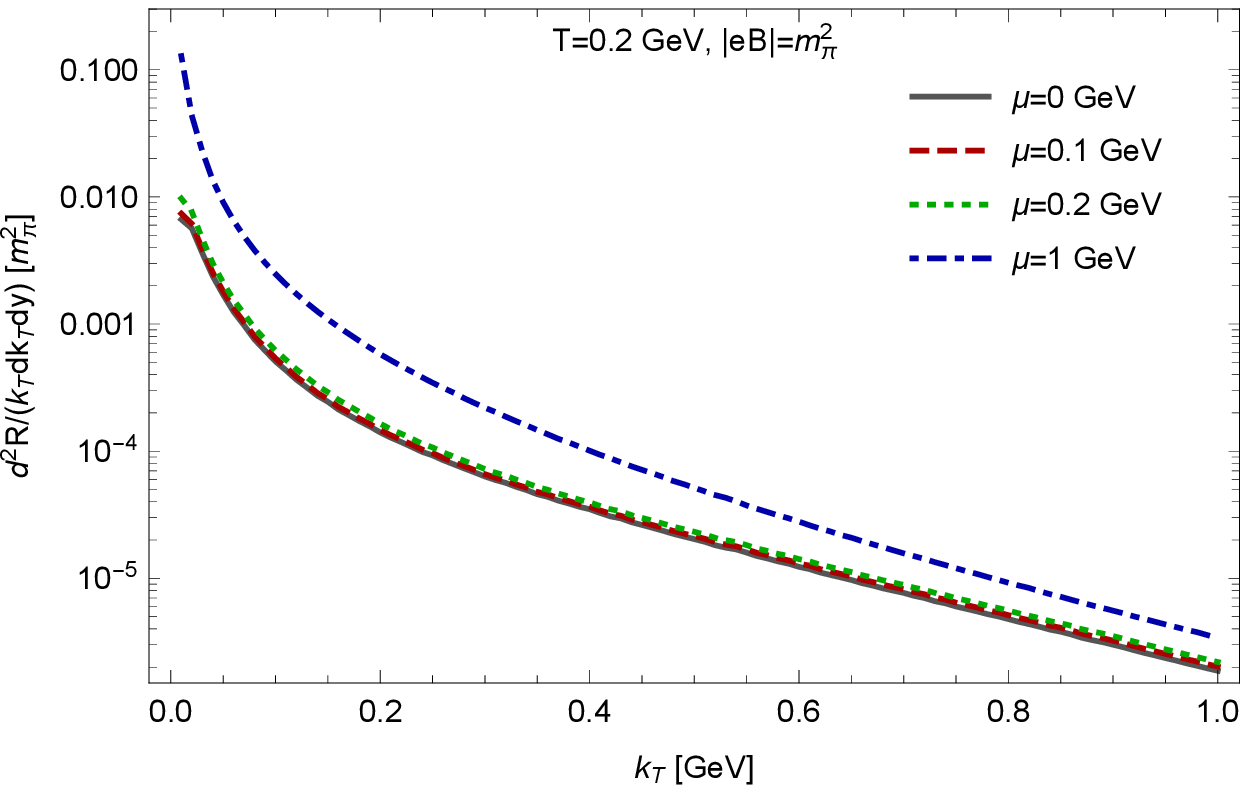}}
\hspace{0.01\textwidth}
\subfigure[]{\includegraphics[width=0.45\textwidth]{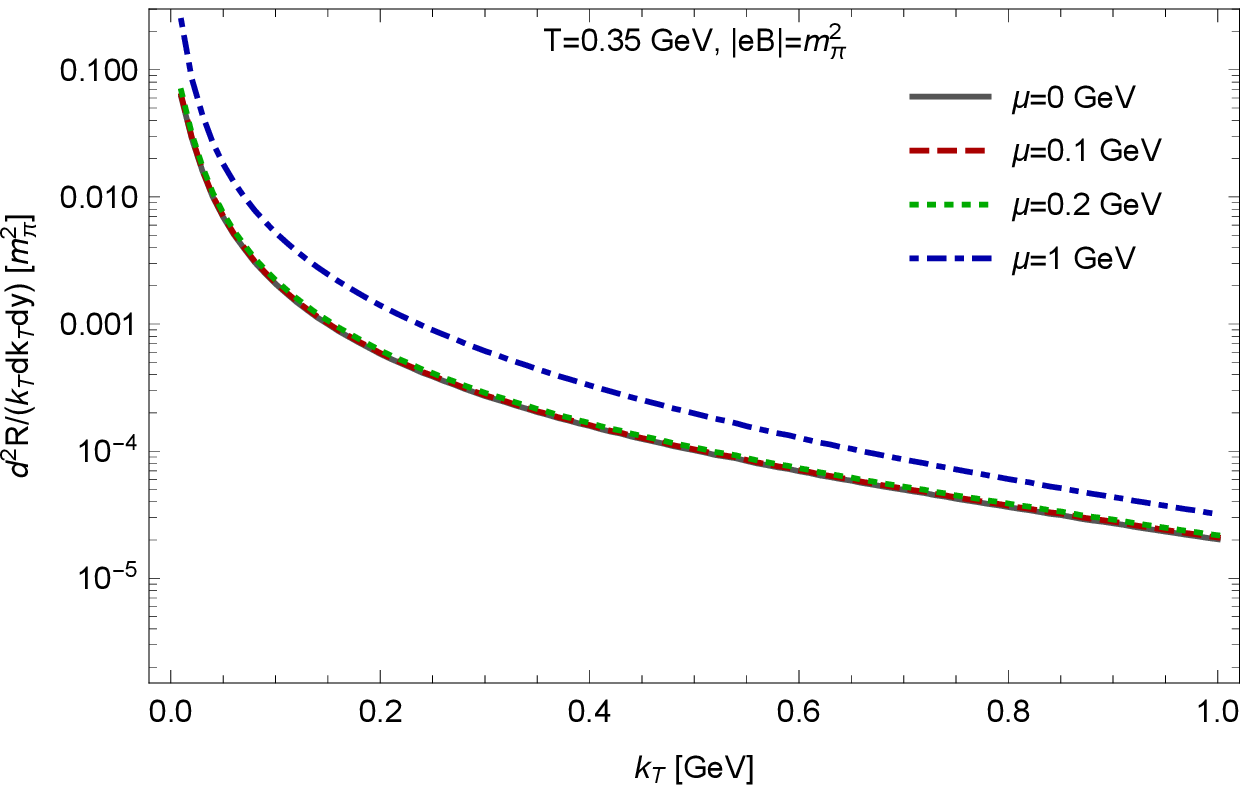}}
\subfigure[]{\includegraphics[width=0.45\textwidth]{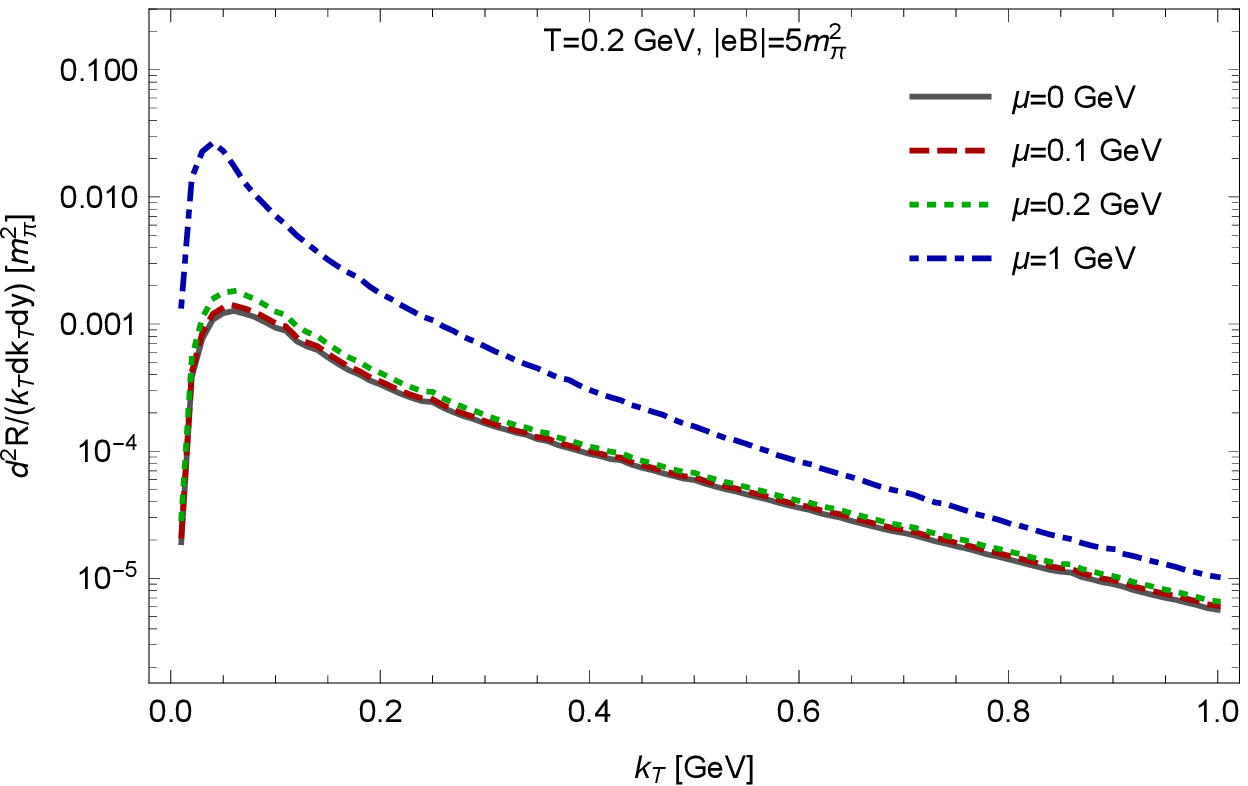}}
\hspace{0.01\textwidth}
\subfigure[]{\includegraphics[width=0.45\textwidth]{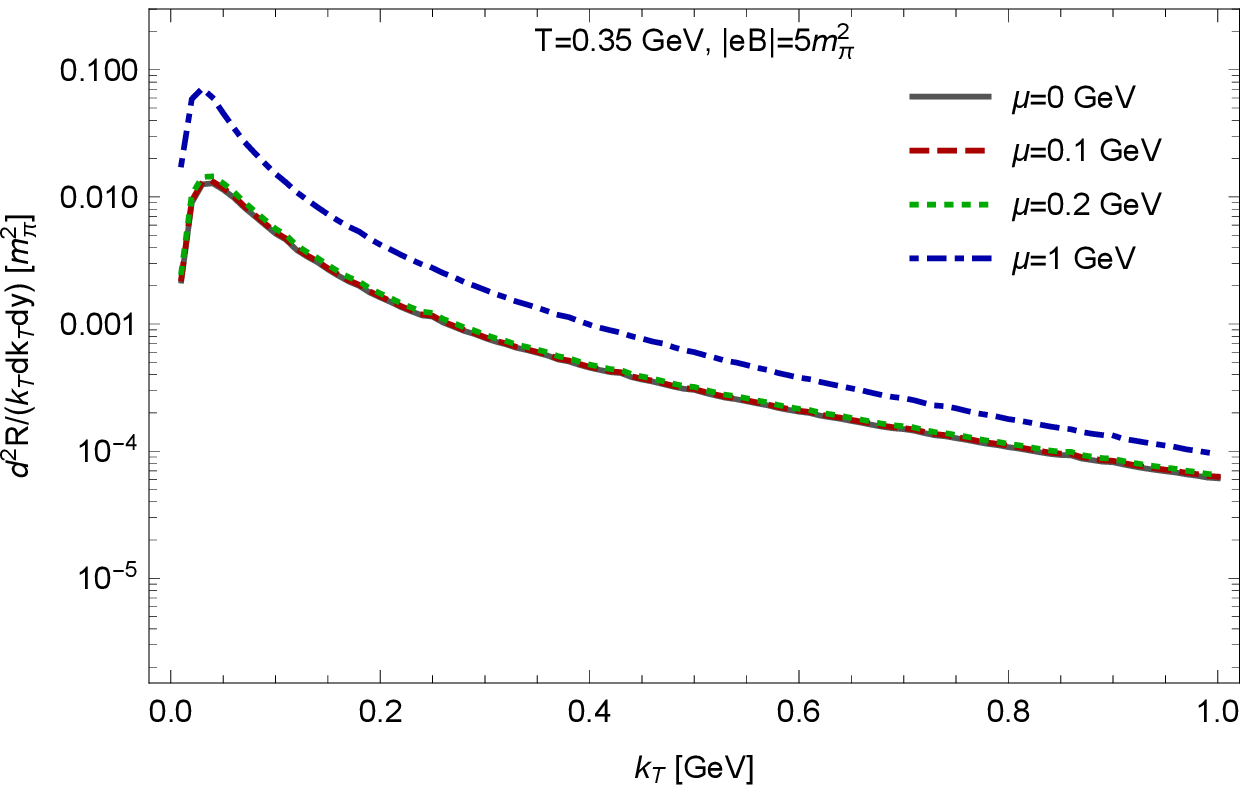}}
\caption{Photon production rate as a function of the transverse momentum $k_T$ for four different values of the chemical potential: $\mu = 0$ (gray lines), $\mu = 0.1~\mbox{GeV}$ (red dashed lines), $\mu = 0.2~\mbox{GeV}$ (green dotted lines), and $\mu = 1~\mbox{GeV}$ (blue dot-dashed lines). The left panels (a and c) show the results for $T=200~\mbox{MeV}$ and the right panels (b and d) for $T=350~\mbox{MeV}$. The top and bottom panels show the results for $|eB| = m_{\pi}^2$ and $|eB| = 5 m_{\pi}^2$, respectively.}
\label{figure-total-rates-B1}
\end{figure}

To understand qualitative effects of a nonzero baryon chemical potential on the photon emission rate, we will start by comparing the results for $\mu = 0$, $\mu = 0.1~\mbox{GeV}$, $\mu = 0.2~\mbox{GeV}$, and $\mu = 1~\mbox{GeV}$, see Fig.~\ref{figure-total-rates-B1}. (Note that the rates at $\mu = 0$ are the same as those reported in Ref.~\cite{Wang:2020dsr} but given in units of $m_{\pi}^2$.) The two smallest values of the chemical potential ($\mu = 0.1~\mbox{GeV}$ and $\mu = 0.2~\mbox{GeV}$) can be viewed as typical for the quark-gluon plasma produced in heavy-ion collisions. While the largest value ($\mu = 1~\mbox{GeV}$) is unrealistic, it is included for instructive purposes to get a deeper insight into the role of the chemical potential under extreme conditions. 

We study the same range of the transverse momenta, from $k_{T,{\rm min}}=0.01~\mbox{GeV}$ to $k_{T,{\rm max}}=1~\mbox{GeV}$, and use the same discretization step $\Delta k_{T}=0.01~\mbox{GeV}$ as in Ref.~\cite{Wang:2020dsr}. Similarly, we cover the same azimuthal angles between $\phi_{\rm min}=10^{-4}\frac{\pi}{2}$ and $\phi_{\rm max}=\frac{\pi}{2} - \phi_{\rm min}$ with the discretization step $\Delta \phi=10^{-3}\frac{\pi}{2}$. To avoid potential problems in numerical calculations, we do not consider the limiting values $\phi=0$ and $\phi=\frac{\pi}{2}$. When evaluating the Landau-level sums, we include a finite but rather large number of Landau levels, i.e., $n_{\rm max} =1000$. Such a choice insures that numerical results are reliable for a sufficiently wide range of transverse momenta: $\sqrt{|eB|} /\sqrt{2n_{\rm max}} \lesssim k_{T} \lesssim \sqrt{2n_{\rm max}|eB|} $.

By comparing the results in Fig.~\ref{figure-total-rates-B1}, we see that the total photon emission rate grows with increasing both temperature and chemical potential. However, the dependence on the chemical potential remains relatively weak for $\mu\lesssim  0.2~\mbox{GeV}$. At its peak value, for example, the rate at $\mu= 0.2~\mbox{GeV}$ is only about 40\% larger than at $\mu=0$. The differences are even smaller away from the peak. In hindsight, this is not surprising since both representative values of temperature are relatively large. With that said, a more careful analysis reveals some surprises. As we will discuss below, each of the partial contributions of the three different types of processes depends much stronger on $\mu$. 

As explained in detail in the earlier studies~\cite{Wang:2020dsr,Wang:2021ebh}, the photon rate must have a local maximum as a function of the transverse momentum (or energy). It is connected with the Landau-level quantization, which becomes important at small $k_{T}$. As we see from Fig.~\ref{figure-total-rates-B1}, a similar peak exists at sufficiently small values of $k_{T}$ also when $\mu\neq 0$. Moreover, the location of the peak does not change much when $\mu\lesssim  0.2~\mbox{GeV}$. At large $\mu$, the maximum tends to shift to smaller values of $k_{T}$. Such a behavior is not surprising since the Landau-level quantization is not affected directly by the chemical potential. However, since a nonzero $\mu$ also changes the occupation numbers of the Landau levels and, in turn, the kinematics of the relevant processes, a weak dependence does appear.

As we stated before, the baryon chemical potential breaks the charge conjugation symmetry. Among other things, this implies that the partial contributions of the quark splitting and antiquark splitting processes (i.e, $q\to q+\gamma$ and $\bar{q}\to \bar{q}+\gamma$, respectively) should be different. For example, when $\mu$ is positive, the relevant number densities of quarks (antiquarks) will be enhanced (suppressed) by the Fermi distribution functions. The corresponding enhancement (suppression) will be also reflected in the photon emission rates. One may expect that the rate of the annihilation process $q+\bar{q}\to \gamma$ is affected as well. 

\begin{figure}[t]
\centering
\subfigure[]{\includegraphics[width=0.45\textwidth]{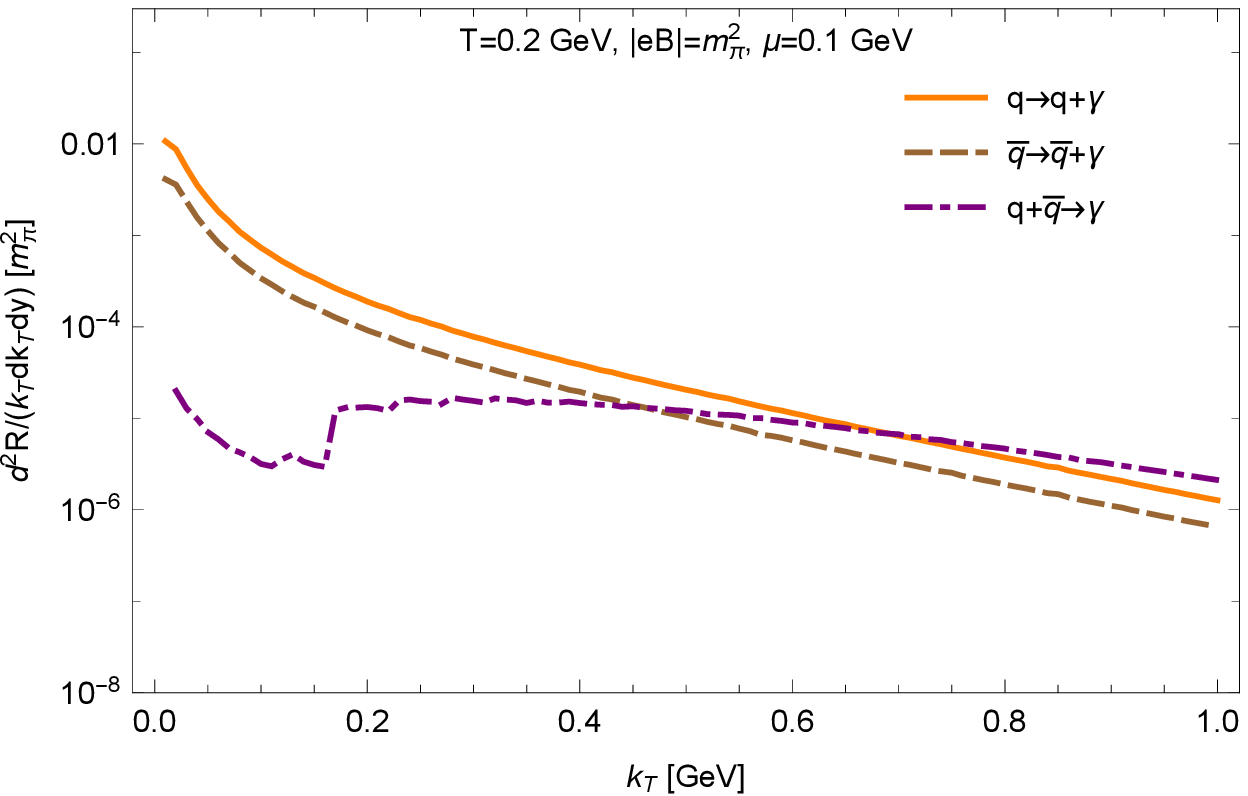}}
\hspace{0.01\textwidth}
\subfigure[]{\includegraphics[width=0.45\textwidth]{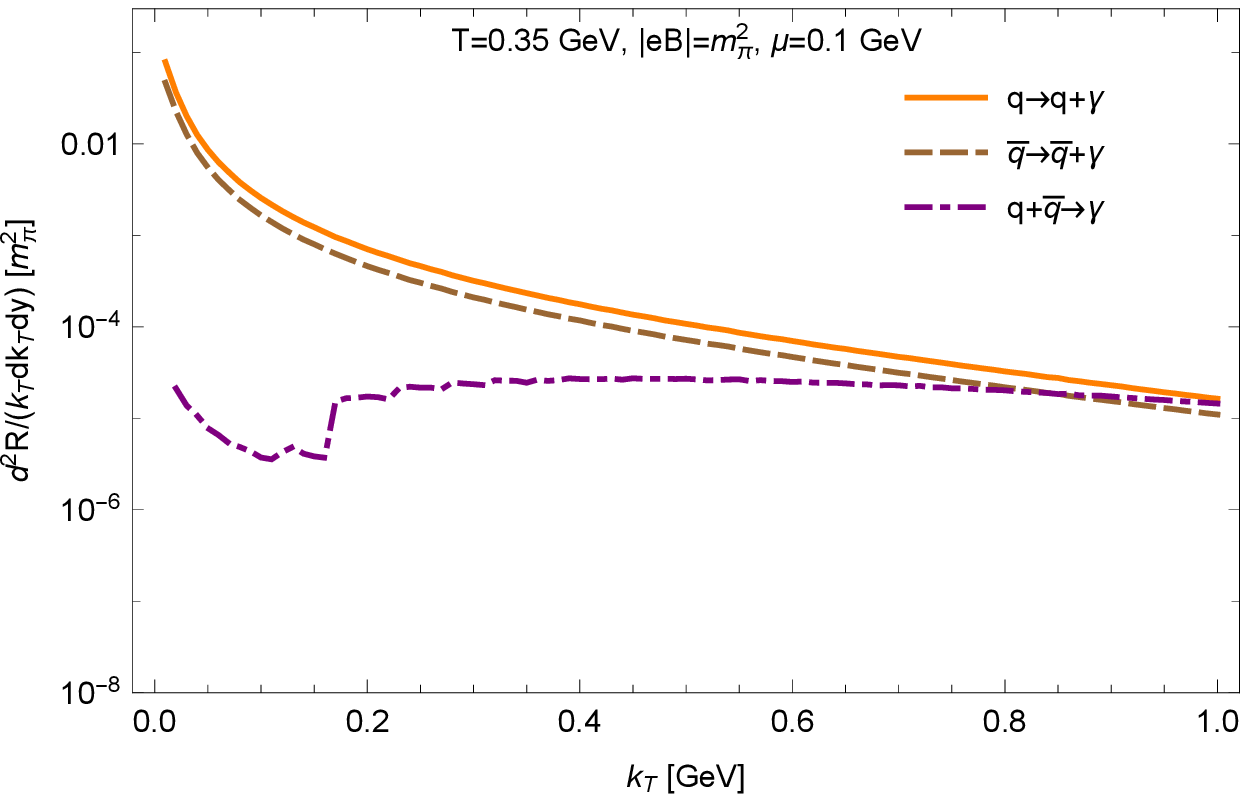}}
\hspace{0.01\textwidth}
\subfigure[]{\includegraphics[width=0.45\textwidth]{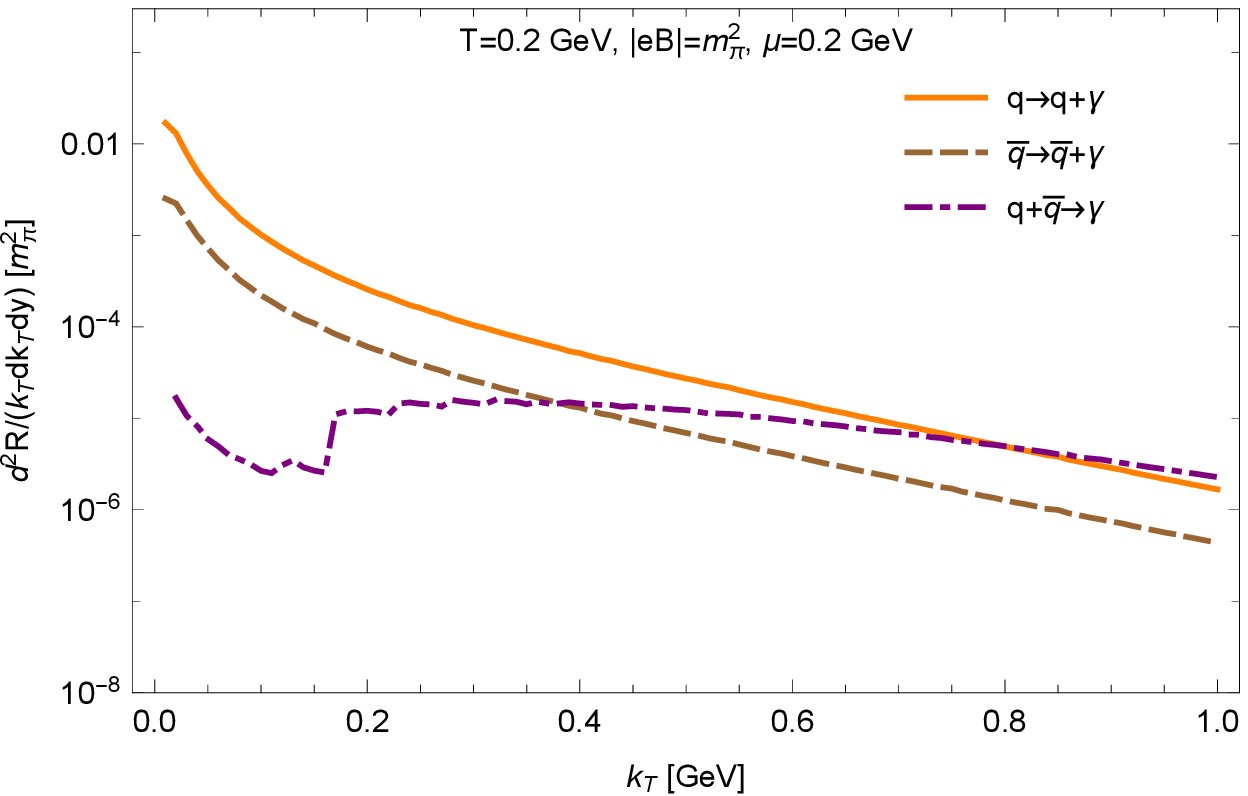}}
\hspace{0.01\textwidth}
\subfigure[]{\includegraphics[width=0.45\textwidth]{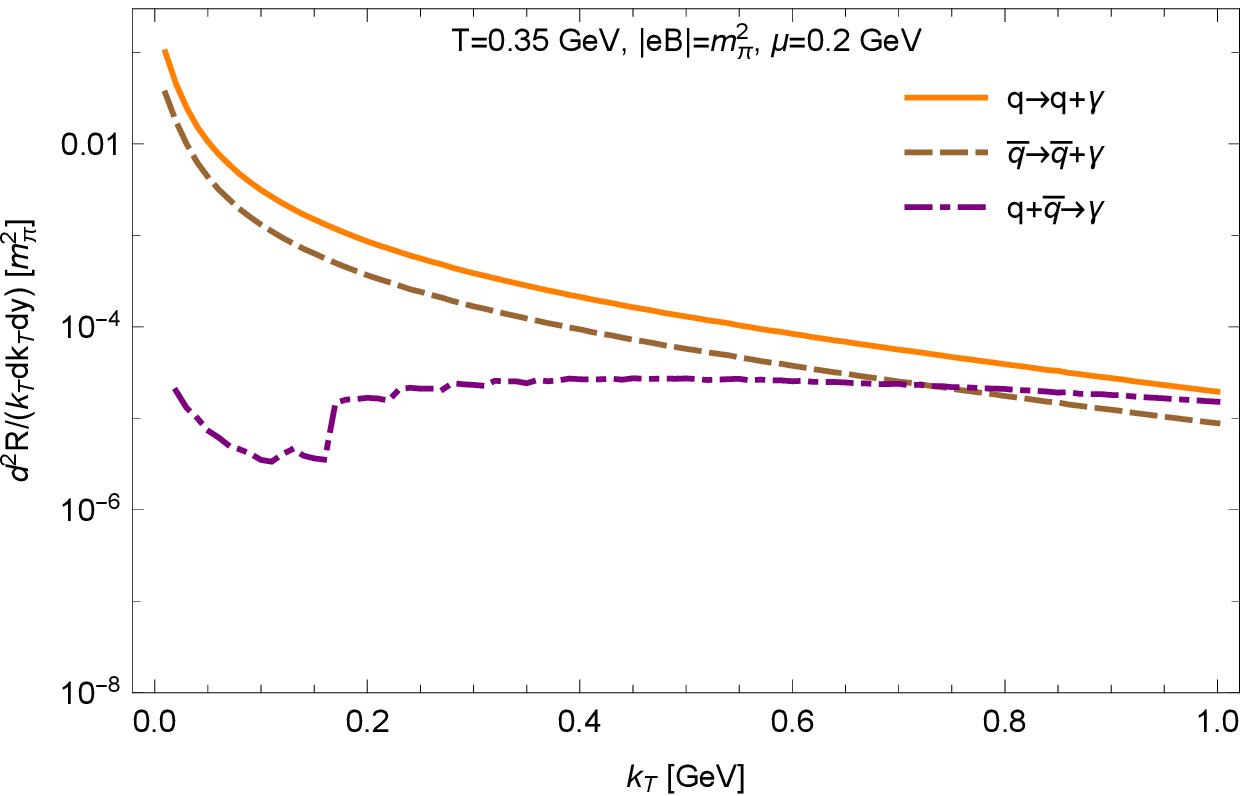}}
\hspace{0.01\textwidth}
\subfigure[]{\includegraphics[width=0.45\textwidth]{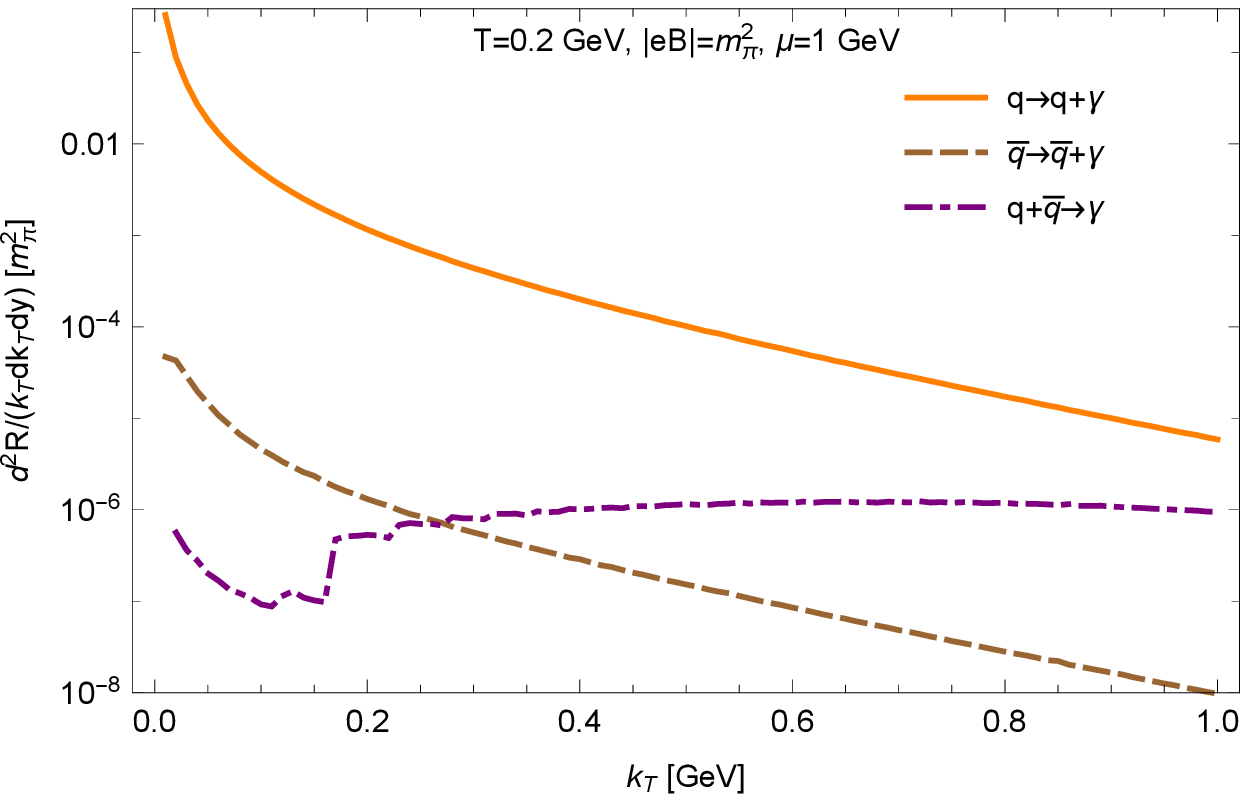}}
\hspace{0.01\textwidth}
\subfigure[]{\includegraphics[width=0.45\textwidth]{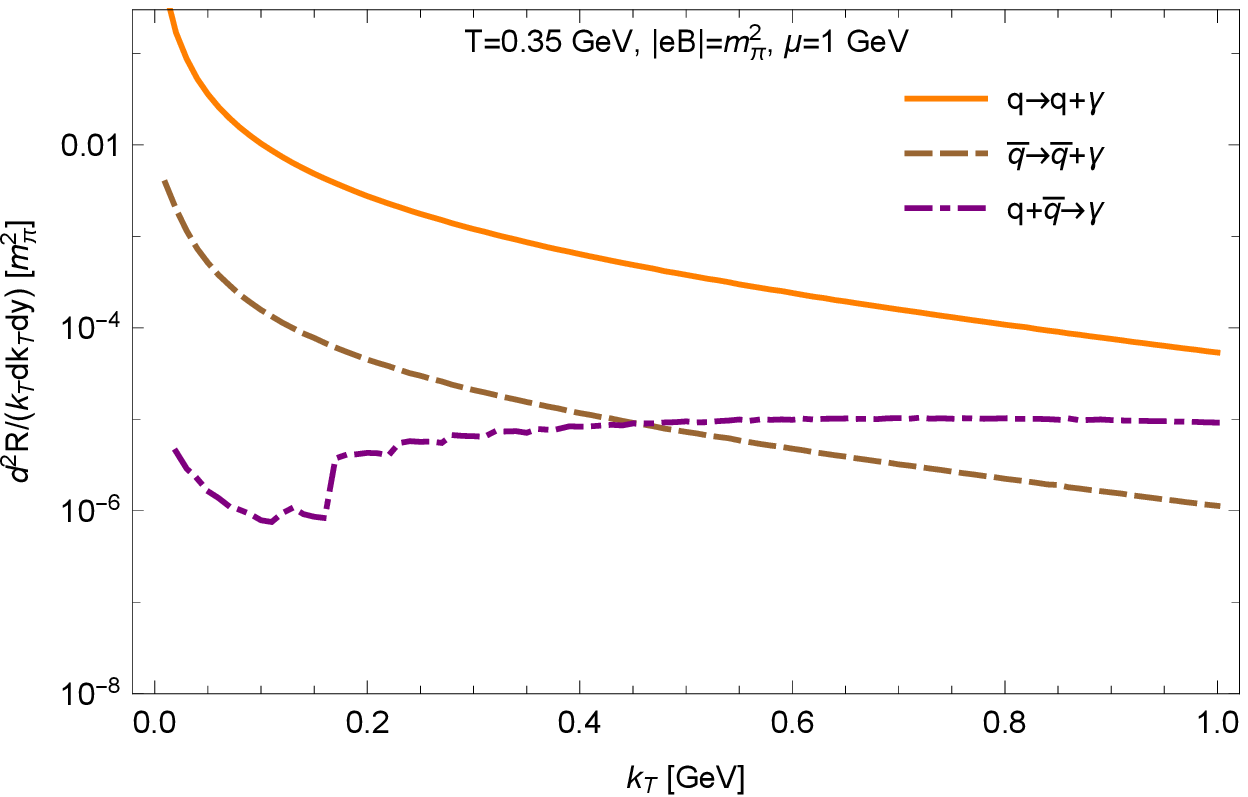}}
\caption{Partial contributions of different types of processes, i.e., $q\to q+\gamma$ (orange lines), $\bar{q}\to \bar{q}+\gamma$ (brown dashed lines), and $q+\bar{q}\to\gamma$ (purple dot-dashed lines), to the photon production rate at $|eB| =  m_{\pi}^2$ as a function of the transverse momentum $k_T$. The left panels (a, c, and e) show the results for $T = 200~\mbox{MeV}$ and the right panels (b, d, and f) for $T =350~\mbox{MeV}$. The top row (panels a and b) shows the results for $\mu = 0.1~\mbox{GeV}$, the middle row (panels a and b) for $\mu = 0.2~\mbox{GeV}$, and the bottom row (panels a and b) for $\mu = 1~\mbox{GeV}$.}
\label{figure-partial-rates-B1}
\end{figure}

In the case of the weaker magnetic field, $|eB| = m_{\pi}^2$, the breakdown of the total photon emission rate into its partial contributions from the three different types of processes is shown in Fig.~\ref{figure-partial-rates-B1}. The results for $T = 200~\mbox{MeV}$  are shown in the three panels on the left (a, c, and e), and the results for $T = 350~\mbox{MeV}$ are shown in the three panels on the right (b, d, and f). It is not surprising that the rates are larger at higher temperature. By comparing the rates at $\mu = 0.1~\mbox{GeV}$ (panels a and b), $\mu = 0.2~\mbox{GeV}$ (panels c and d), and $\mu = 1~\mbox{GeV}$ (panels e and f), we observe a qualitative dependence that was expected from general considerations. First, the difference between the rates of the two processes, $q\to q+\gamma$ and $\bar{q}\to \bar{q}+\gamma$, grows with $\mu$. Second, the corresponding difference grows faster and becomes more pronounced at $T = 200~\mbox{MeV}$, compared to the case of $T = 350~\mbox{MeV}$. Again, this is not surprising since a growing temperature tends to wash away the effects of a nonzero $\mu$. Third, the annihilation rate remains relatively small compared to the rate of the quark splitting $q\to q+\gamma$ and, to a lesser degree, even the antiquark splitting $\bar{q}\to \bar{q}+\gamma$. The hierarchiy of rates tends to change at sufficiently large values of $k_{T}$. The switch of the regimes, where the rates of $\bar{q}\to \bar{q}+\gamma$  and $q+\bar{q}\to \gamma$ become equal, is pushed to smaller $k_{T}$ when $\mu$ increases. On the other hand, the switch of the regimes, where the rates of $q\to q+\gamma$ and $\bar{q}\to \bar{q}+\gamma$ become equal, is pushed to higher $k_{T}$.

\begin{figure}[t]
\centering
\subfigure[]{\includegraphics[width=0.45\textwidth]{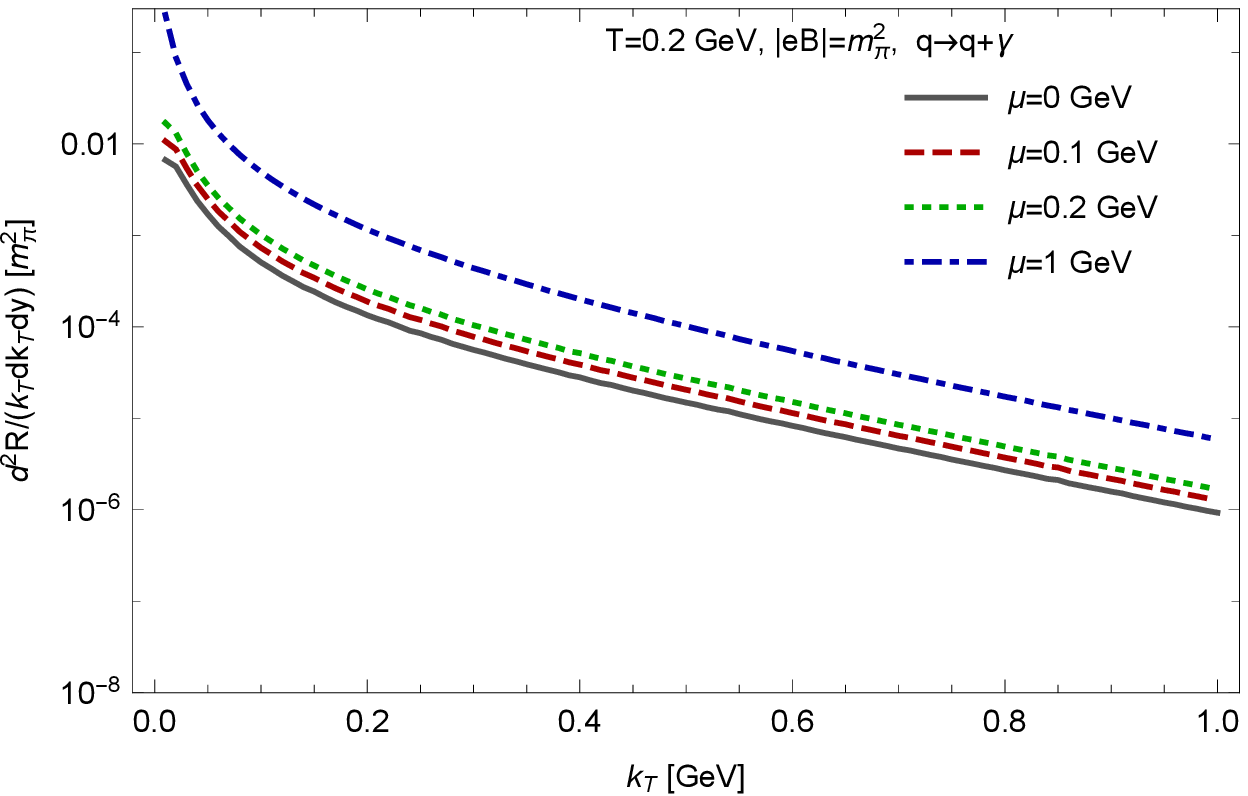}}
\hspace{0.01\textwidth}
\subfigure[]{\includegraphics[width=0.45\textwidth]{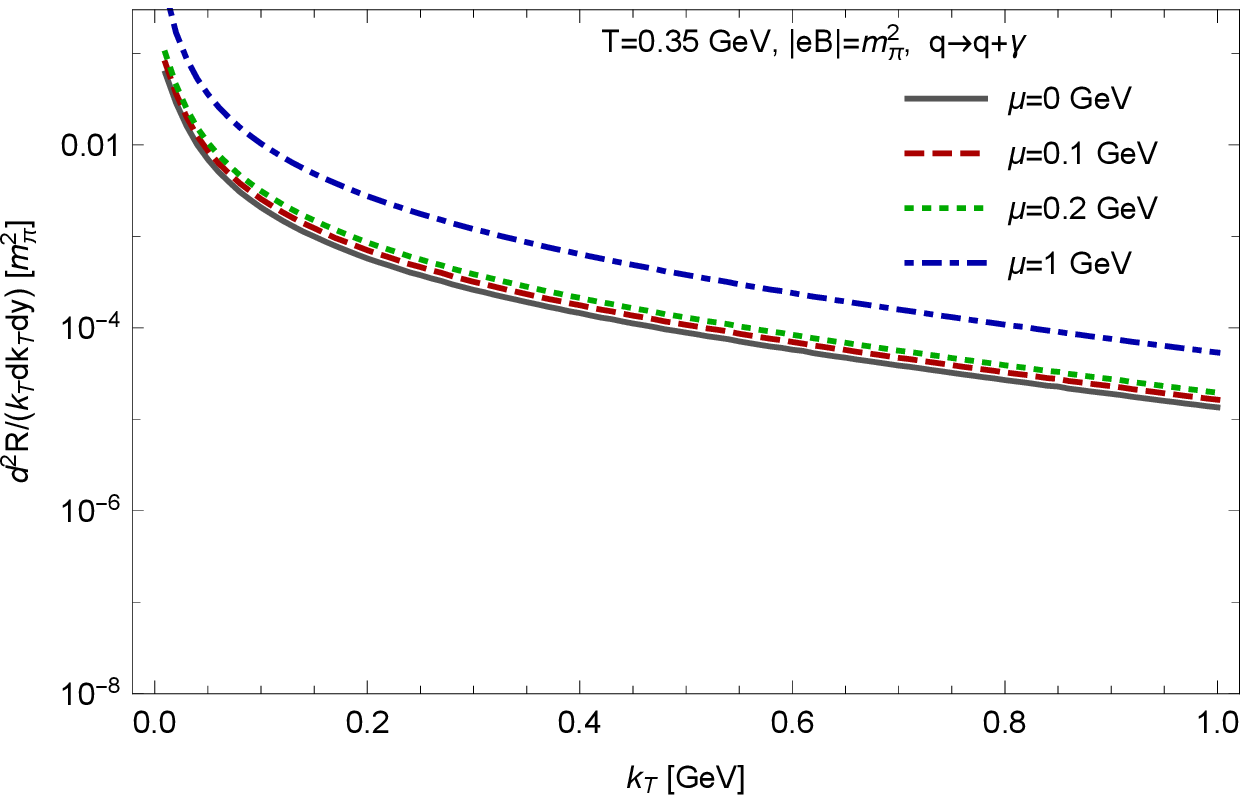}}
\hspace{0.01\textwidth}
\subfigure[]{\includegraphics[width=0.45\textwidth]{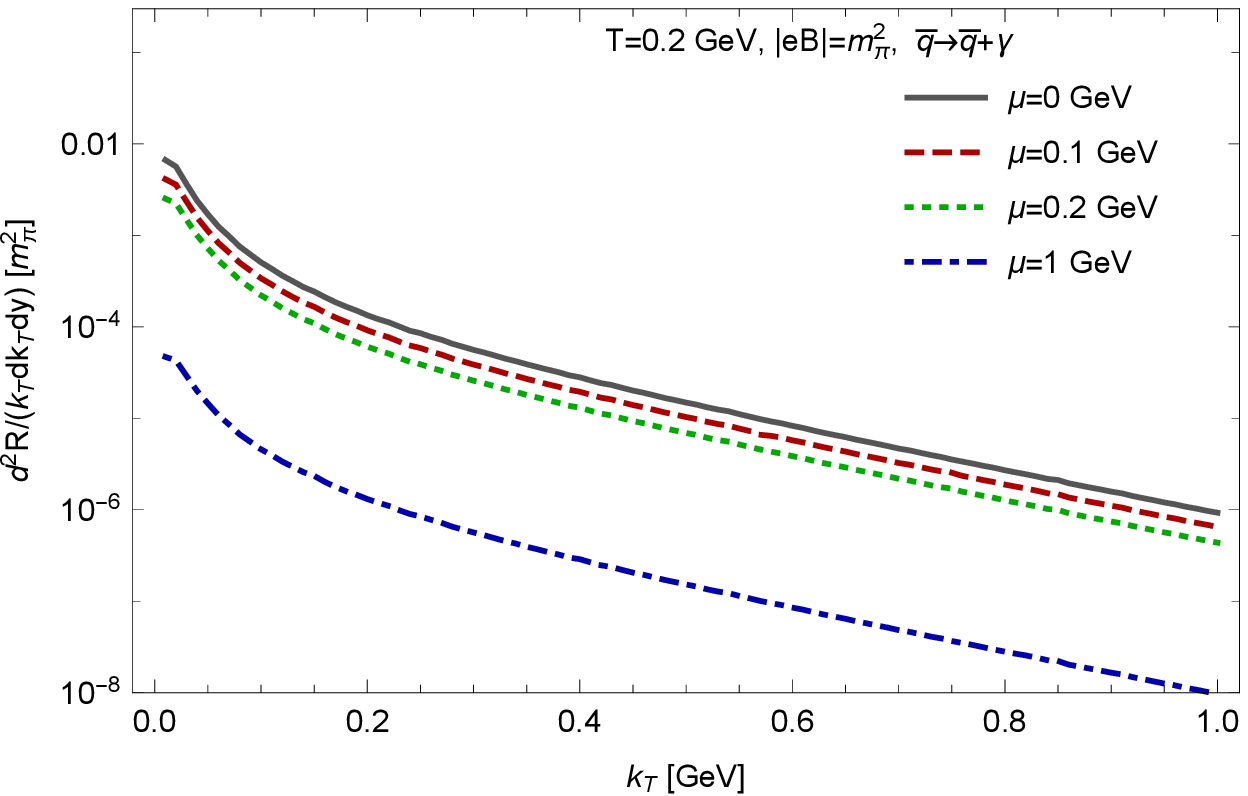}}
\hspace{0.01\textwidth}
\subfigure[]{\includegraphics[width=0.45\textwidth]{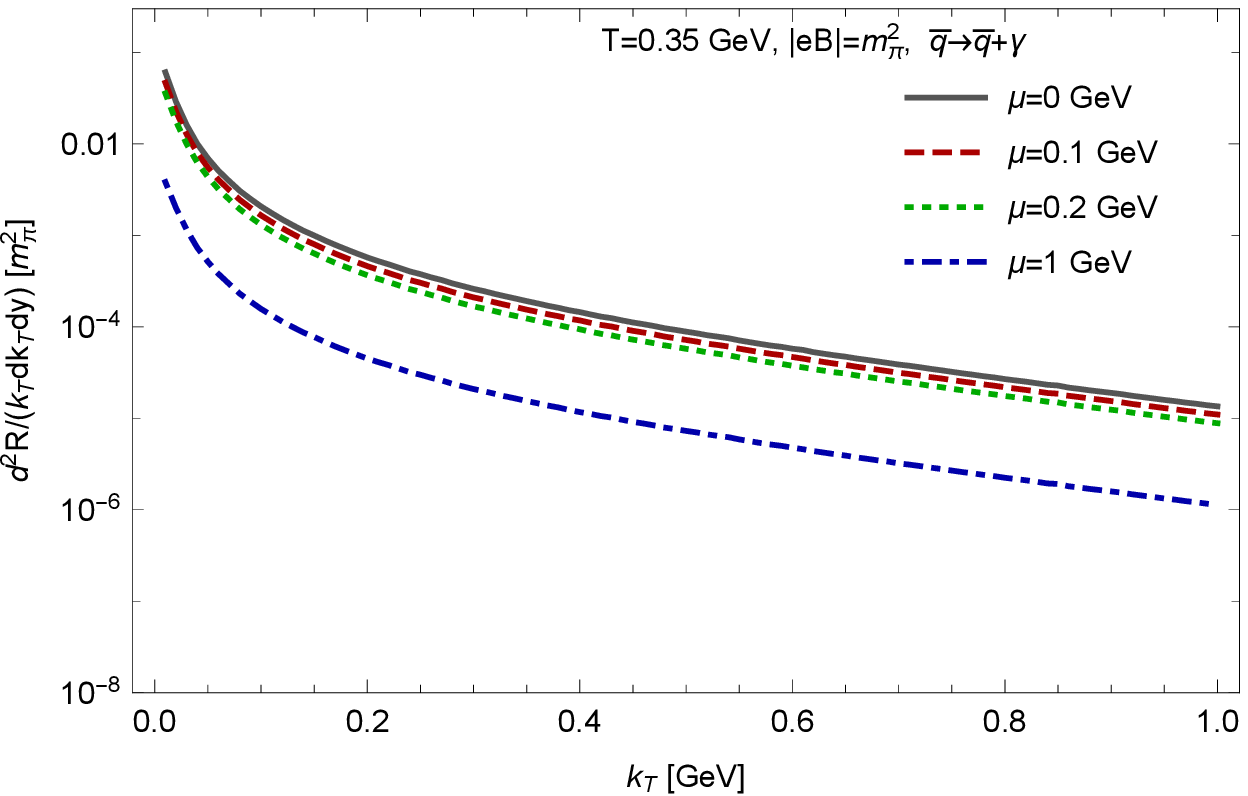}}
\hspace{0.01\textwidth}
\subfigure[]{\includegraphics[width=0.45\textwidth]{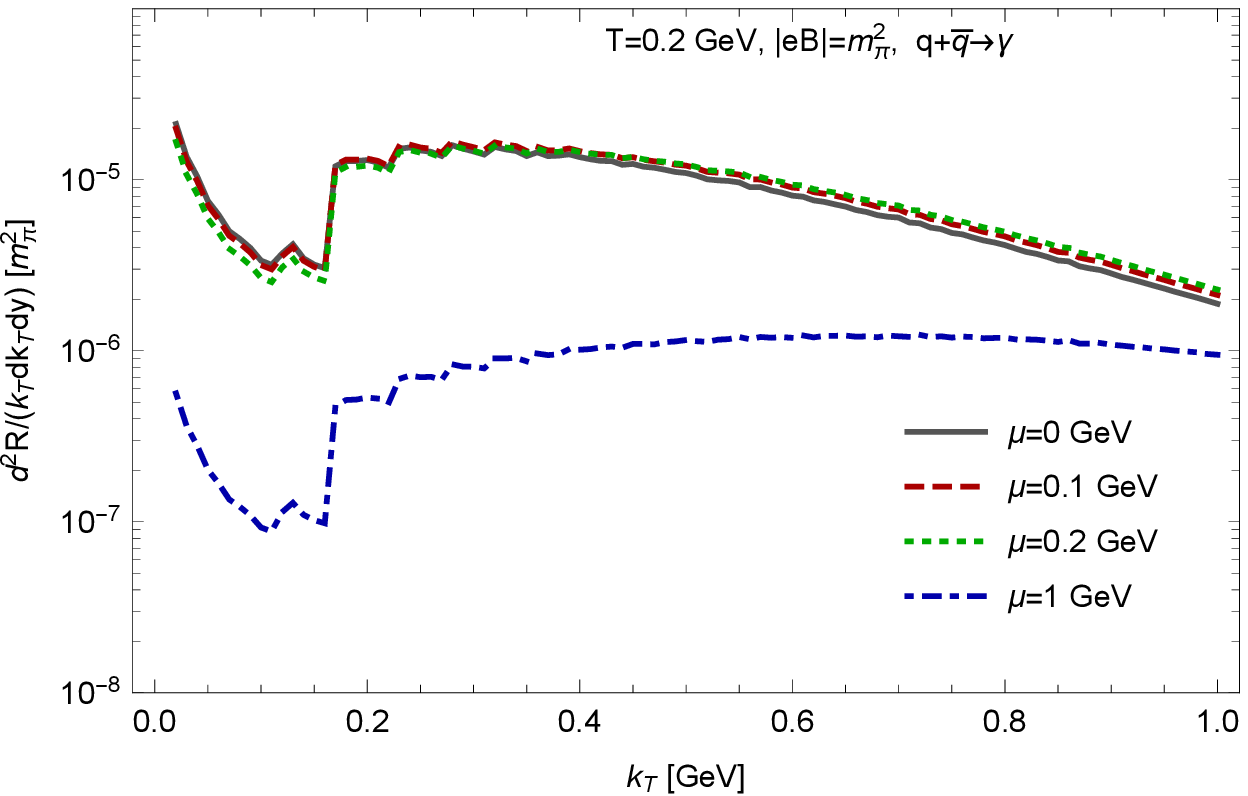}}
\hspace{0.01\textwidth}
\subfigure[]{\includegraphics[width=0.45\textwidth]{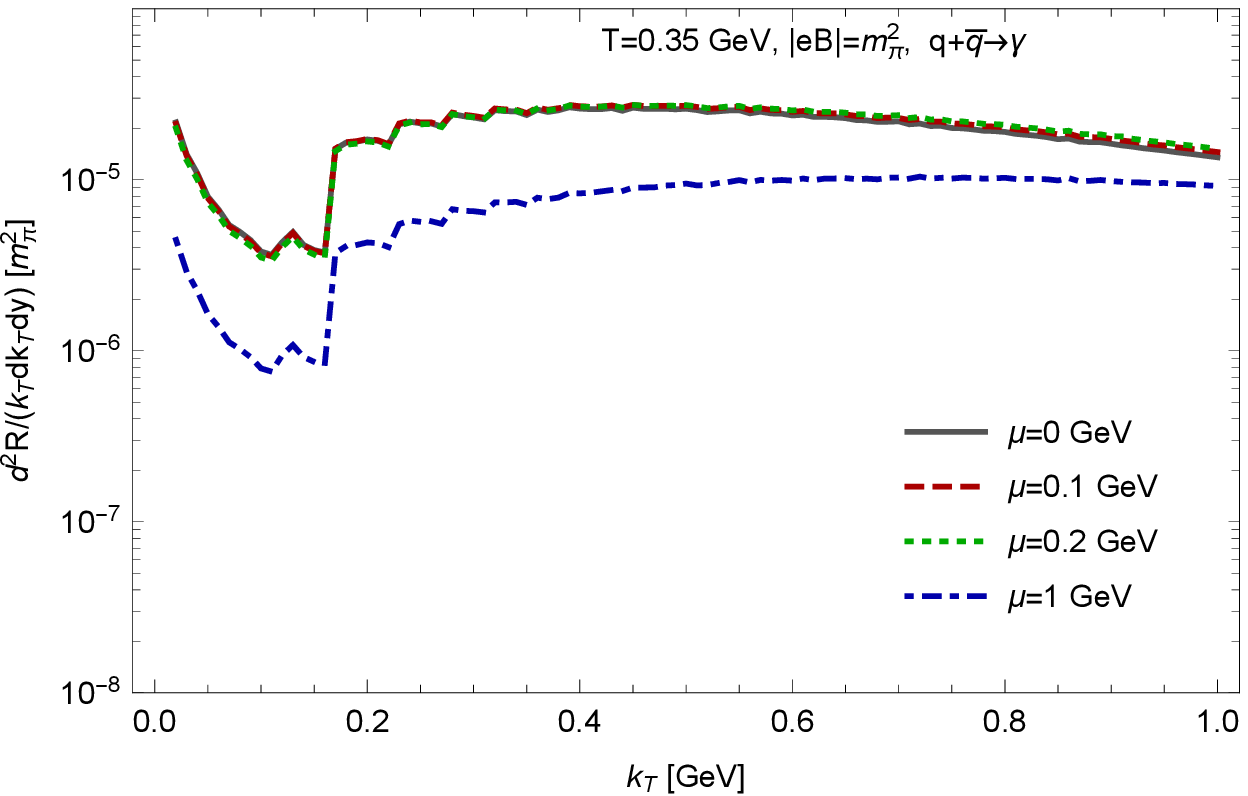}}
\caption{Partial contributions of the three types of processes to the photon production rate at $|eB| =  m_{\pi}^2$ as functions of the transverse momentum $k_T$ for four different values of the chemical potential: $\mu = 0$ (gray lines), $\mu = 0.1~\mbox{GeV}$ (red dashed lines), $\mu = 0.2~\mbox{GeV}$ (green dotted lines), and $\mu = 1~\mbox{GeV}$ (blue dot-dashed lines). The left panels (a, c, and e) show the results for $T = 200~\mbox{MeV}$ and the right panels (b, d, and f) for $T =350~\mbox{MeV}$. The top row (panels a and b) gives the rates due to the quark splitting $q\to q+\gamma$, the middle row (panels a and b) gives the rates due to the antiquark splitting $\bar{q}\to \bar{q}+\gamma$, and the bottom row (panels a and b) gives the rates due to the quark-antiquark annihilation $q+\bar{q}\to \gamma$. Note that the range on the vertical axis is different in the two bottom panels.}
\label{figure-separate-rates-B1}
\end{figure}

To show how the rate for each process type changes with $\mu$, we presented their partial contributions in several panels of Fig.~\ref{figure-separate-rates-B1}. Each panel represents only one of the processes but  combines the results for all four different values of the chemical potential. As we see from panels (a) and (b), the rate of the quark splitting $q\to q+\gamma$ increases with $\mu$ for both values of the temperatures, $T = 200~\mbox{MeV}$ and  $T =350~\mbox{MeV}$. The behavior is opposite for the antiquark splitting $\bar{q}\to \bar{q}+\gamma$, shown in panels (c) and (d). The corresponding rates decrease with $\mu$. The situation for the annihilation rate, represented by panels (e) and (f), is somewhat more complicated. (Note that the range on the vertical axis is different in these two panels.) While the overall rate tends to decrease with $\mu$, it may have a non-monotonous dependence on $\mu$ at large $k_T$ values. 

\begin{figure}[t]
\centering
\subfigure[]{\includegraphics[width=0.45\textwidth]{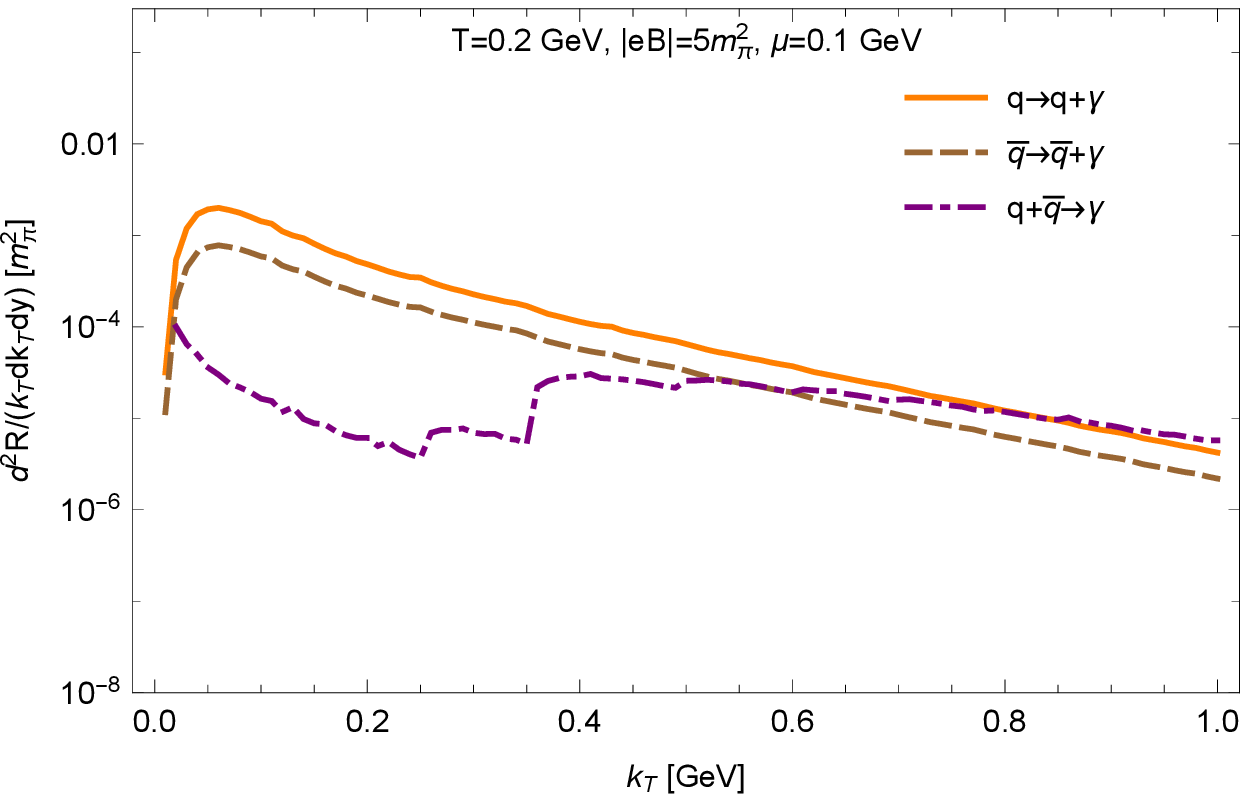}}
\hspace{0.01\textwidth}
\subfigure[]{\includegraphics[width=0.45\textwidth]{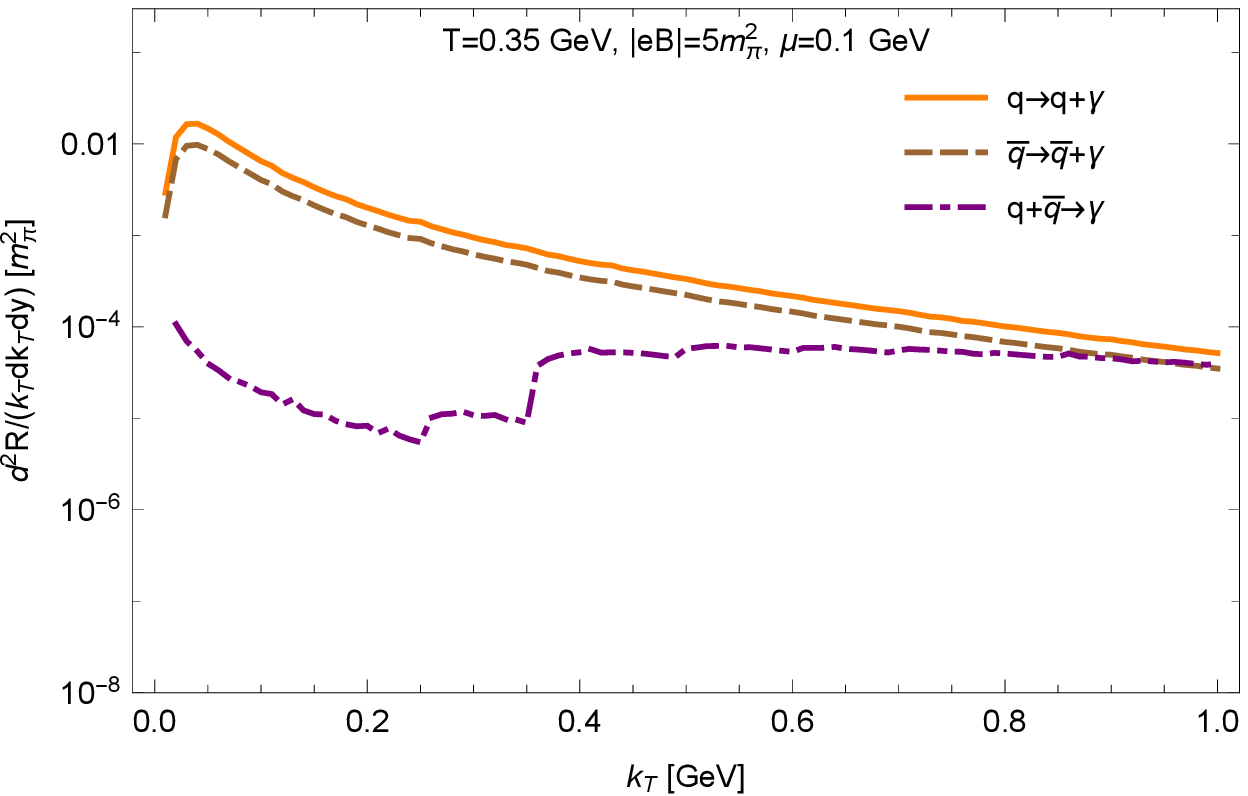}}
\hspace{0.01\textwidth}
\subfigure[]{\includegraphics[width=0.45\textwidth]{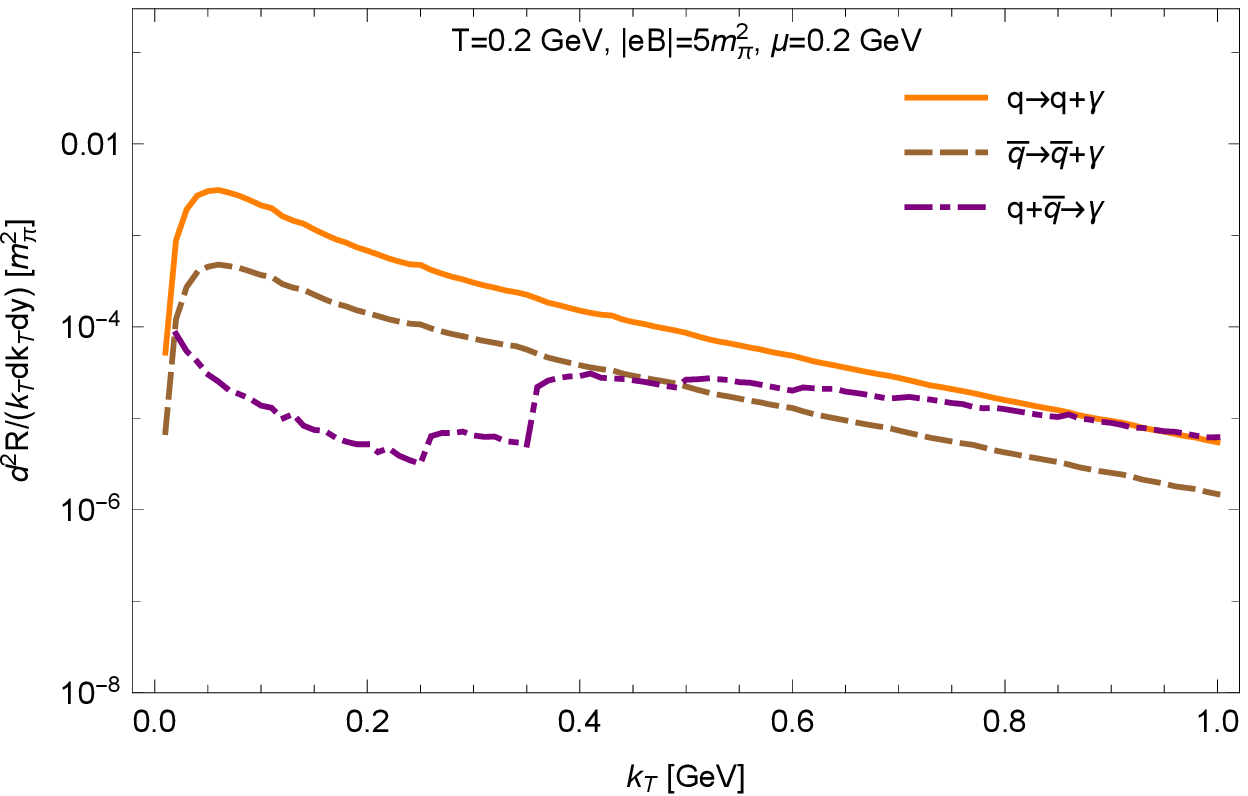}}
\hspace{0.01\textwidth}
\subfigure[]{\includegraphics[width=0.45\textwidth]{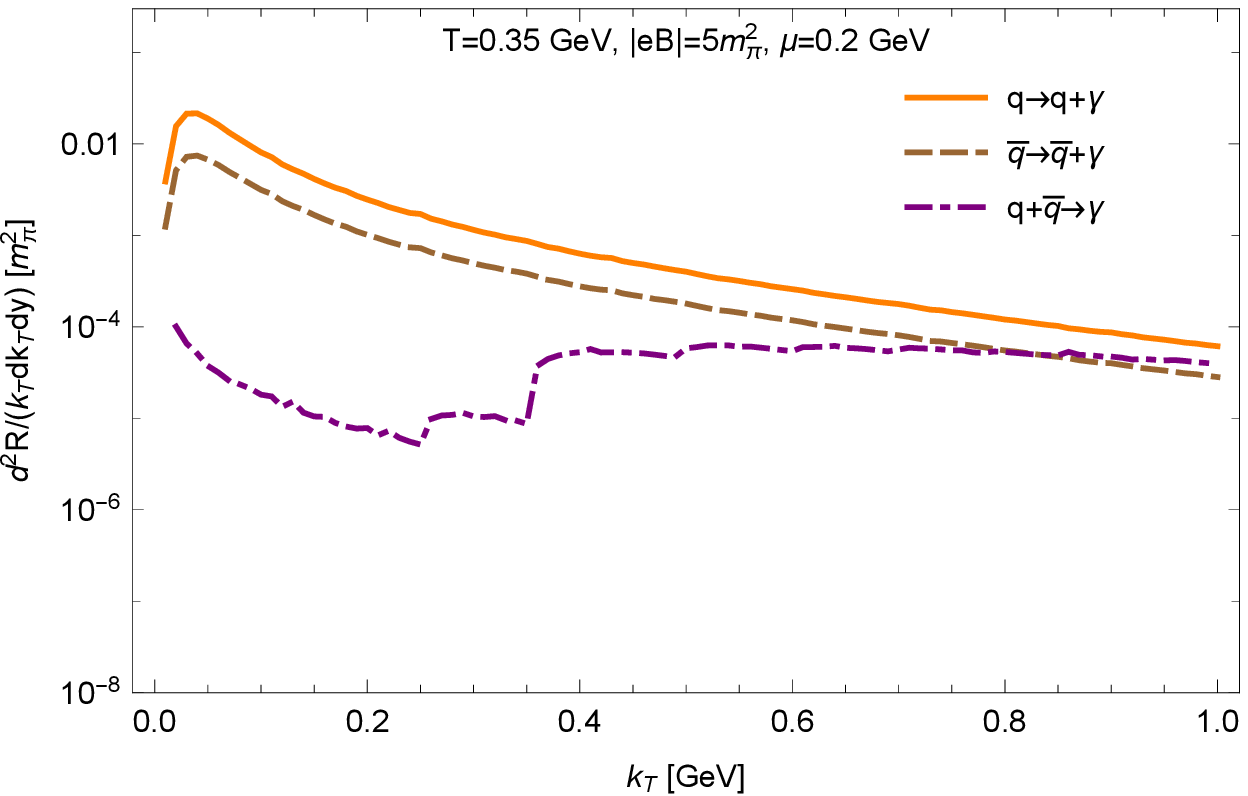}}
\hspace{0.01\textwidth}
\subfigure[]{\includegraphics[width=0.45\textwidth]{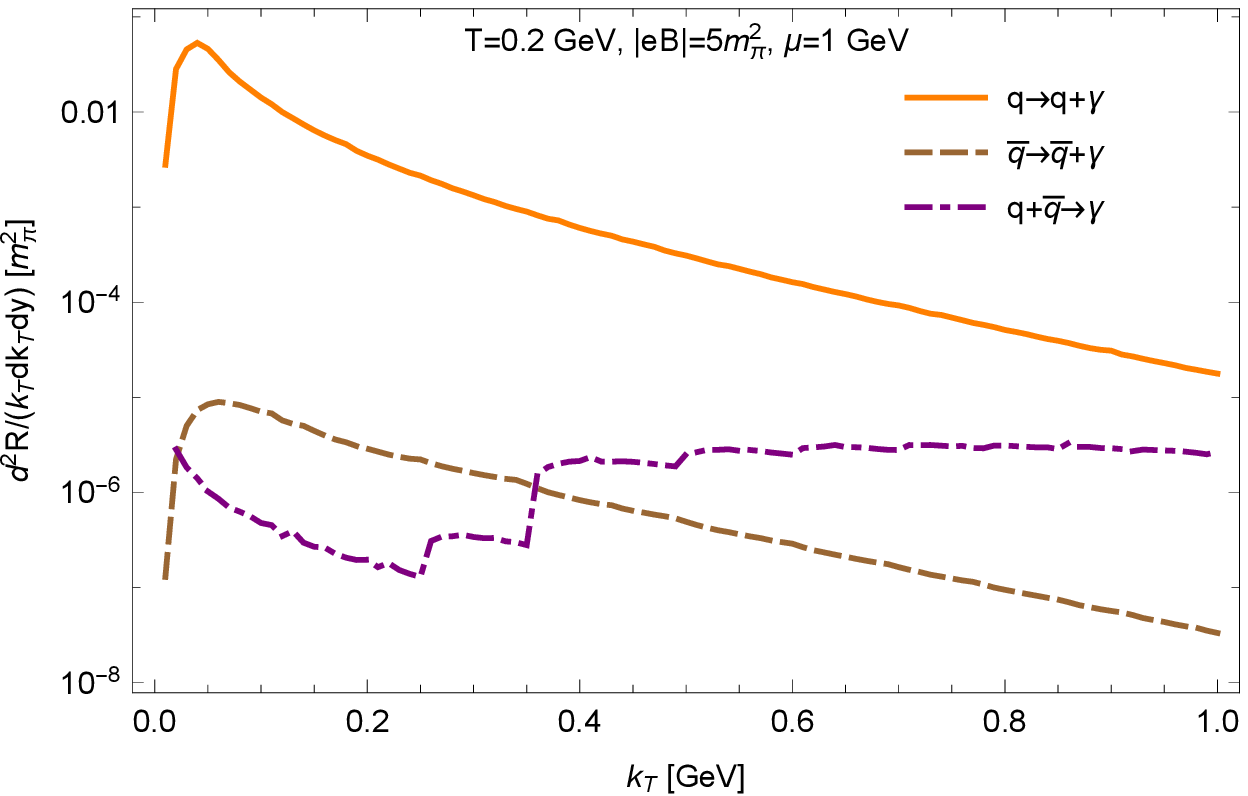}}
\hspace{0.01\textwidth}
\subfigure[]{\includegraphics[width=0.45\textwidth]{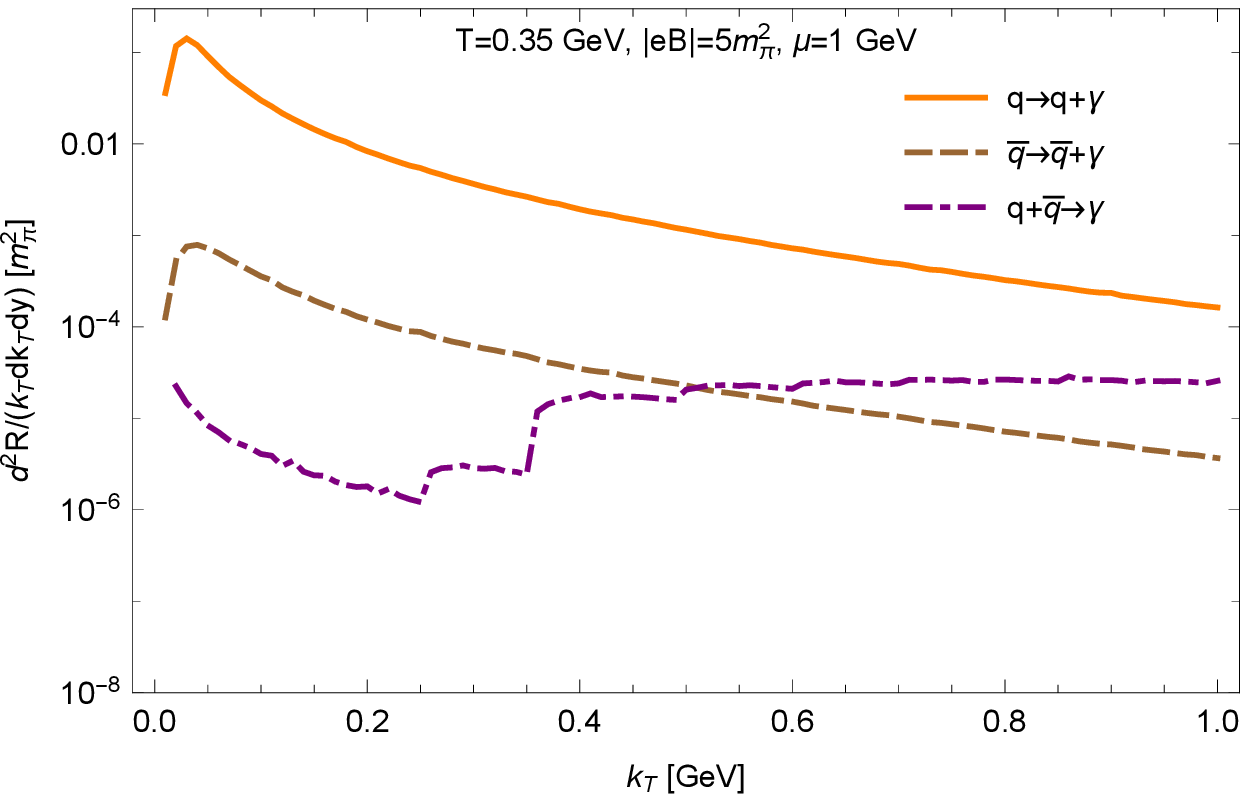}}
\caption{Partial contributions of different types of processes, i.e., $q\to q+\gamma$ (orange lines), $\bar{q}\to \bar{q}+\gamma$ (brown dashed lines), and $q+\bar{q}\to\gamma$ (purple dot-dashed lines), to the photon production rate at $|eB| =  5 m_{\pi}^2$ as a function of the transverse momentum $k_T$. The left panels (a, c, and e) show the results for $T = 200~\mbox{MeV}$ and the right panels (b, d, and f) for $T =350~\mbox{MeV}$. The top row (panels a and b) shows the results for $\mu = 0.1~\mbox{GeV}$, the middle row (panels a and b) for $\mu = 0.2~\mbox{GeV}$, and the bottom row (panels a and b) for $\mu = 1~\mbox{GeV}$.} 
\label{figure-partial-rates-B5}
\end{figure}

In the case of the stronger magnetic field, $|eB| = 5 m_{\pi}^2$, the breakdown of the total photon emission rate into its partial contributions from the three different types of processes is shown in multiple panels of Fig.~\ref{figure-partial-rates-B5}. As we see, all qualitative features remain the same as in the case of the weaker field. In particular, with growing $\mu$, the difference between the rates of quark and antiquark splitting processes increases, but the effect is less pronounced at a higher temperature. The interplay of the annihilation and the two splitting processes is qualitatively the same as in Fig.~\ref{figure-partial-rates-B1}.

\begin{figure}[t]
\centering
\subfigure[]{\includegraphics[width=0.45\textwidth]{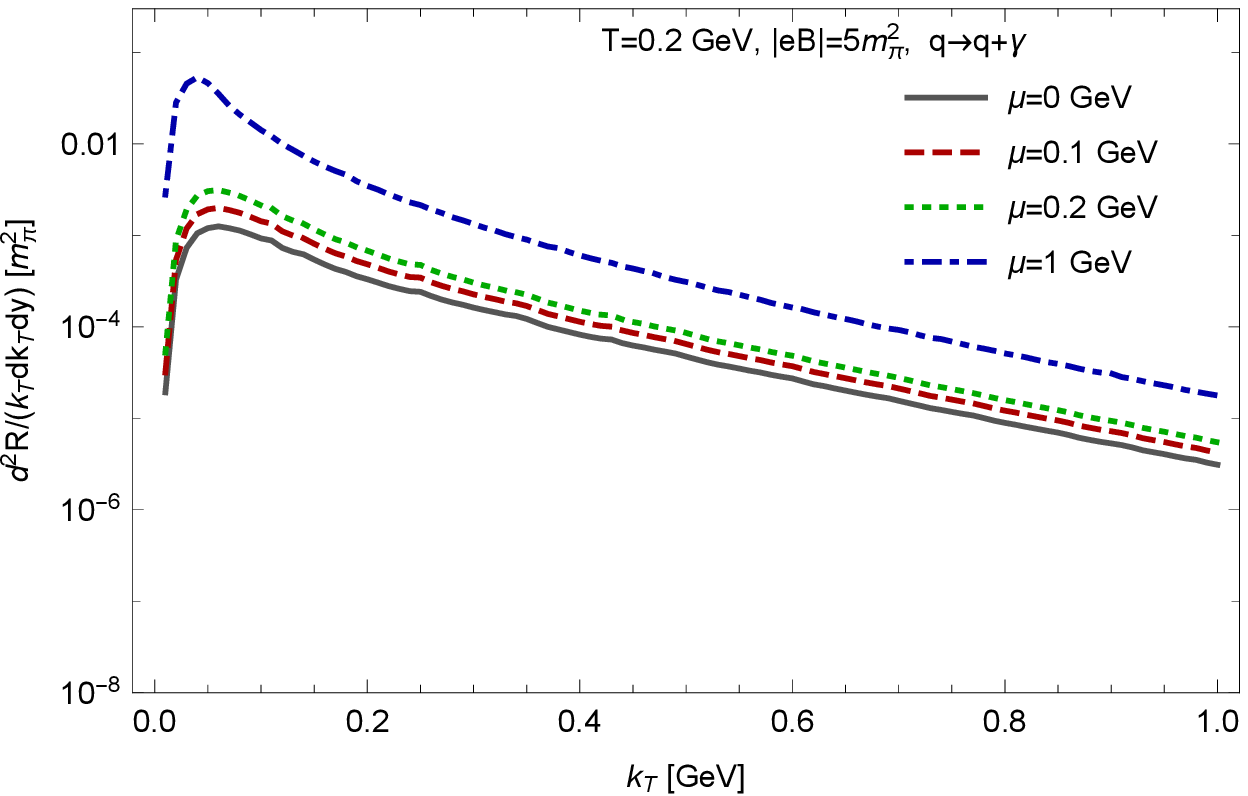}}
\hspace{0.01\textwidth}
\subfigure[]{\includegraphics[width=0.45\textwidth]{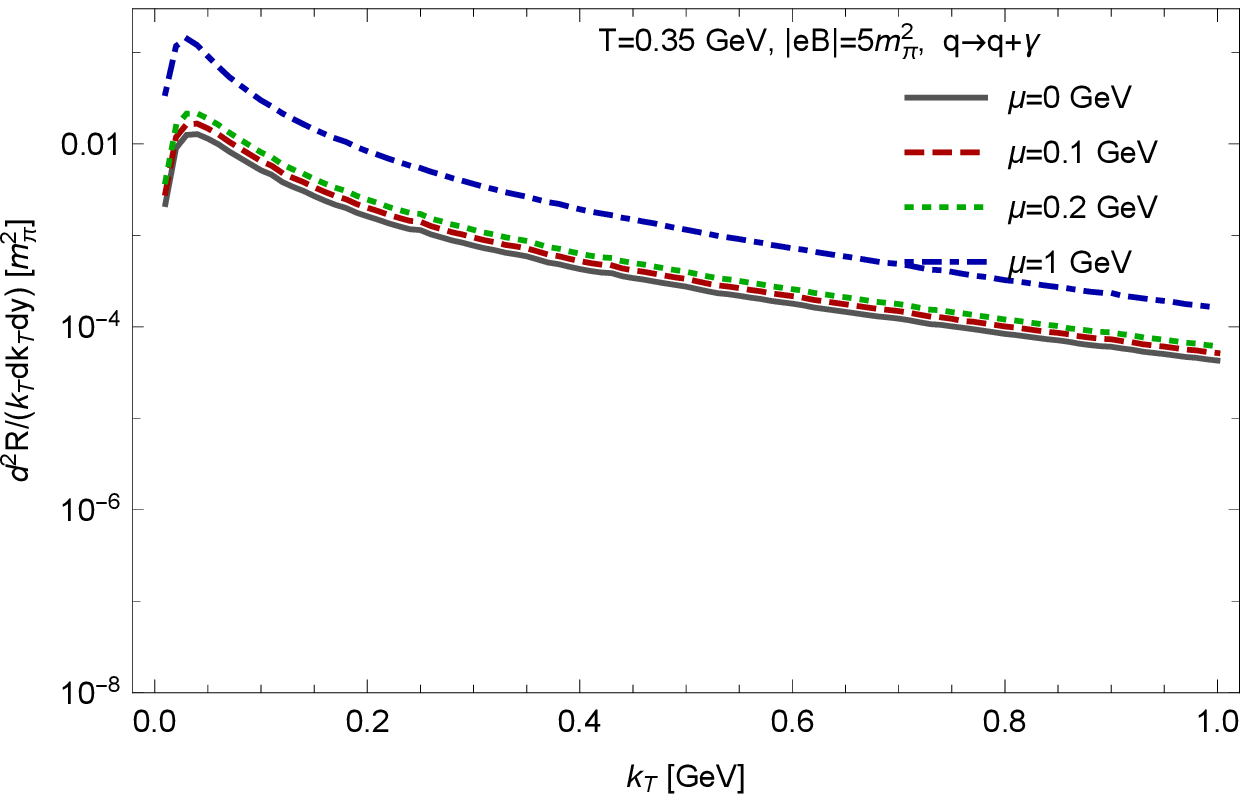}}
\hspace{0.01\textwidth}
\subfigure[]{\includegraphics[width=0.45\textwidth]{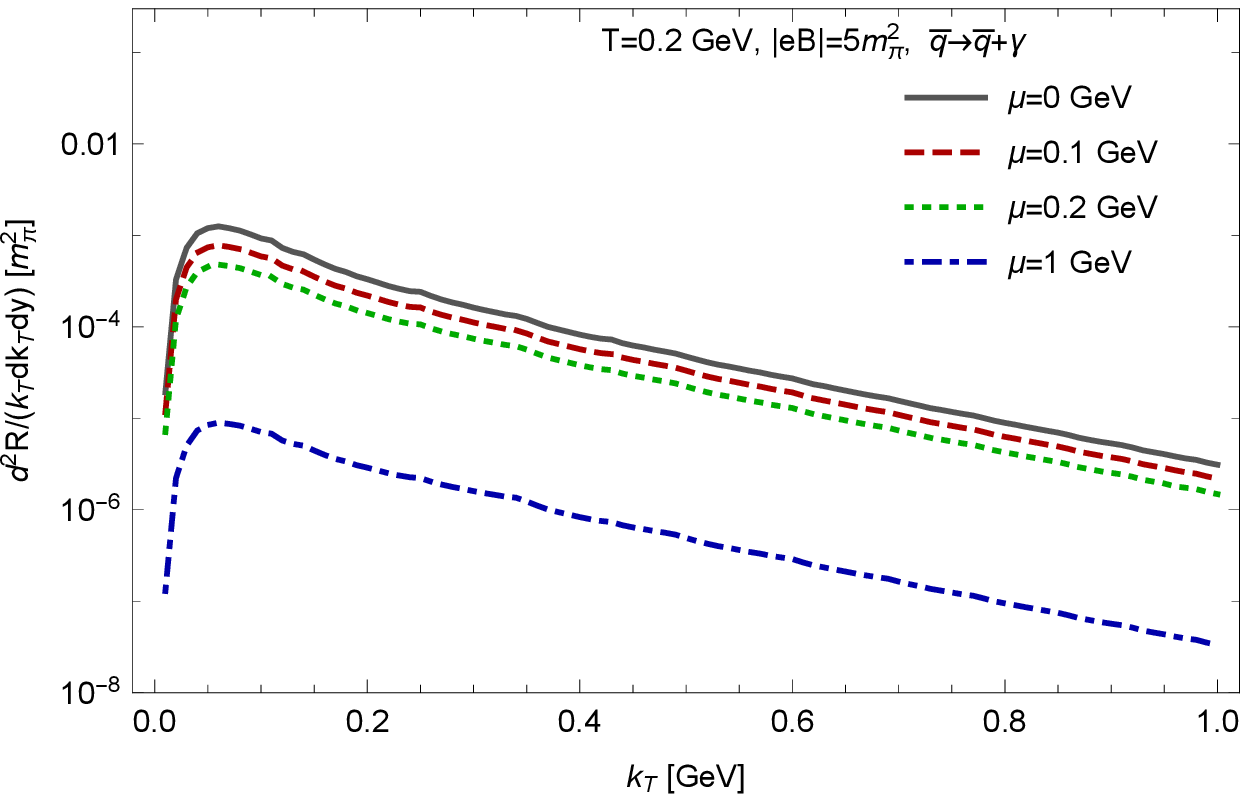}}
\hspace{0.01\textwidth}
\subfigure[]{\includegraphics[width=0.45\textwidth]{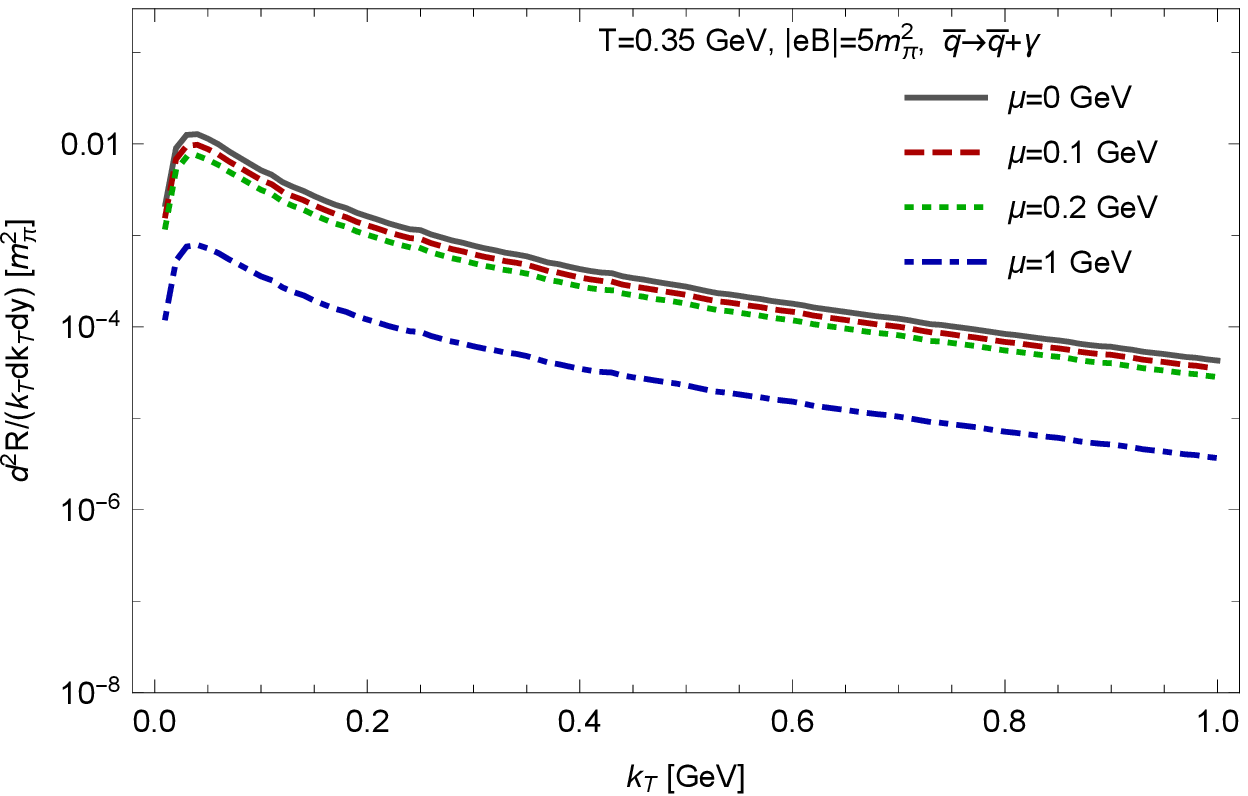}}
\hspace{0.01\textwidth}
\subfigure[]{\includegraphics[width=0.45\textwidth]{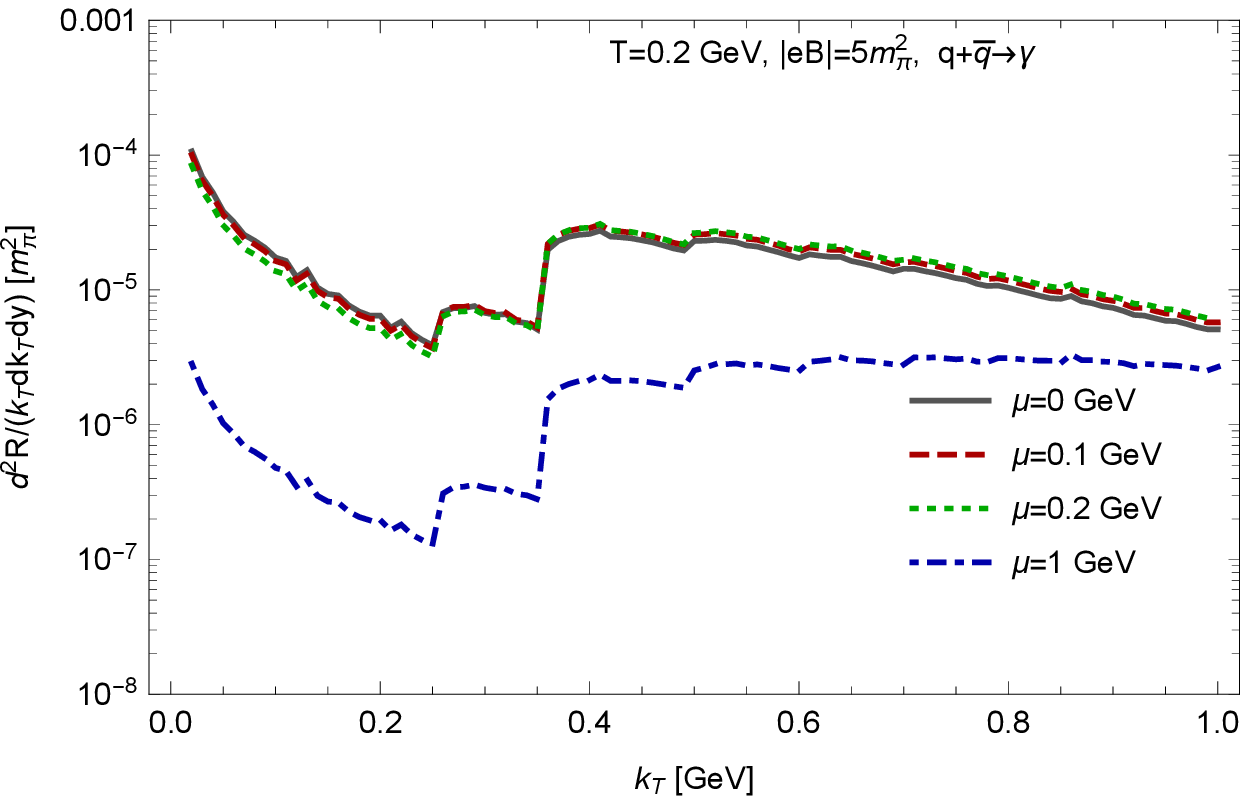}}
\hspace{0.01\textwidth}
\subfigure[]{\includegraphics[width=0.45\textwidth]{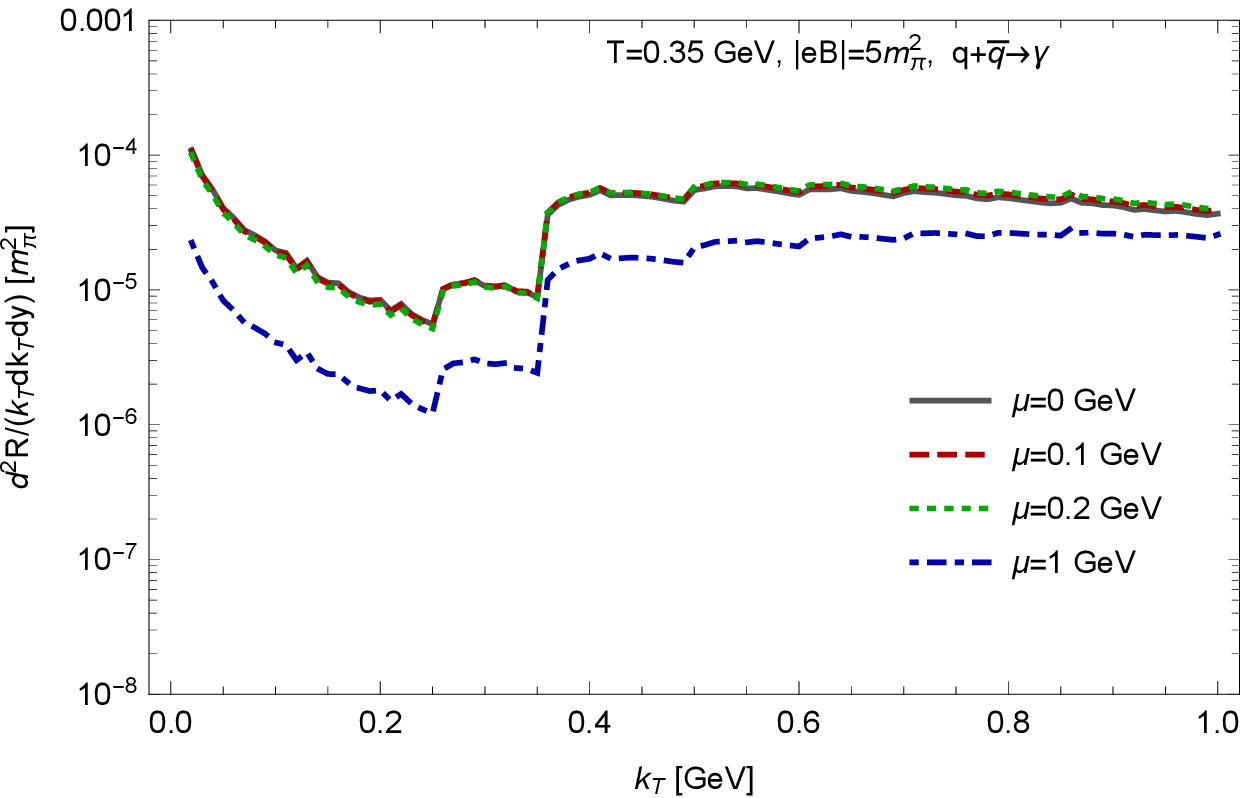}}
\caption{Partial contributions of the three types of processes to the photon production rate at $|eB| =  5m_{\pi}^2$ as functions of the transverse momentum $k_T$ for four different values of the chemical potential: $\mu = 0$ (gray lines), $\mu = 0.1~\mbox{GeV}$ (red dashed lines), $\mu = 0.2~\mbox{GeV}$ (green dotted lines), and $\mu = 1~\mbox{GeV}$ (blue dot-dashed lines). The left panels (a, c, and e) show the results for $T = 200~\mbox{MeV}$ and the right panels (b, d, and f) for $T =350~\mbox{MeV}$. The top row (panels a and b) gives the rates due to the quark splitting $q\to q+\gamma$, the middle row (panels a and b) gives the rates due to the antiquark splitting $\bar{q}\to \bar{q}+\gamma$, and the bottom row (panels a and b) gives the rates due to the quark-antiquark annihilation $q+\bar{q}\to \gamma$. Note that the range on the vertical axis is different in the two bottom panels.}
\label{figure-separate-rates-B5}
\end{figure}

For the $|eB| = 5 m_{\pi}^2$ case, the partial rates of different process types for all four values of the chemical potential are summarized in several panels of Fig.~\ref{figure-separate-rates-B5}. Again, the qualitative features remain the same  as in the case of the weaker  magnetic field in Fig.~\ref{figure-separate-rates-B1}. 

\begin{figure}[t]
\centering
\subfigure[]{\includegraphics[width=0.45\textwidth]{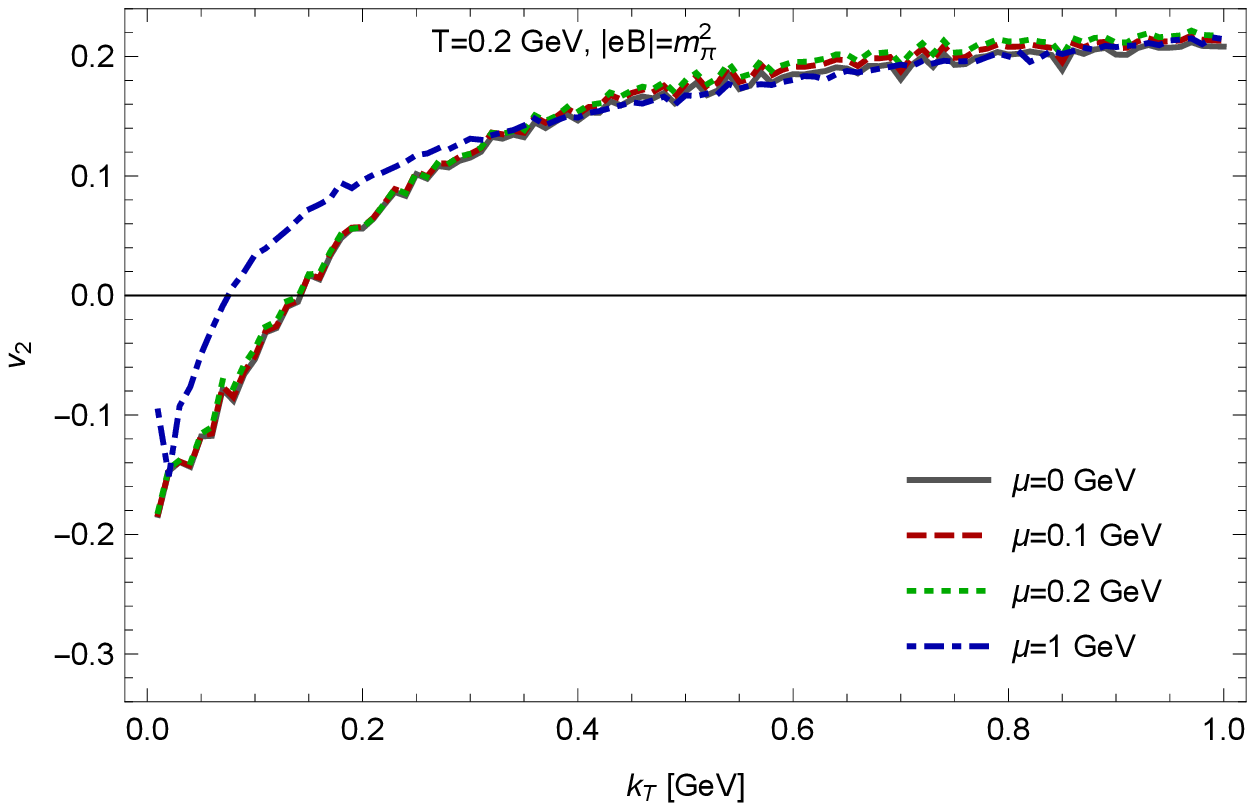}}
\hspace{0.01\textwidth}
\subfigure[]{\includegraphics[width=0.45\textwidth]{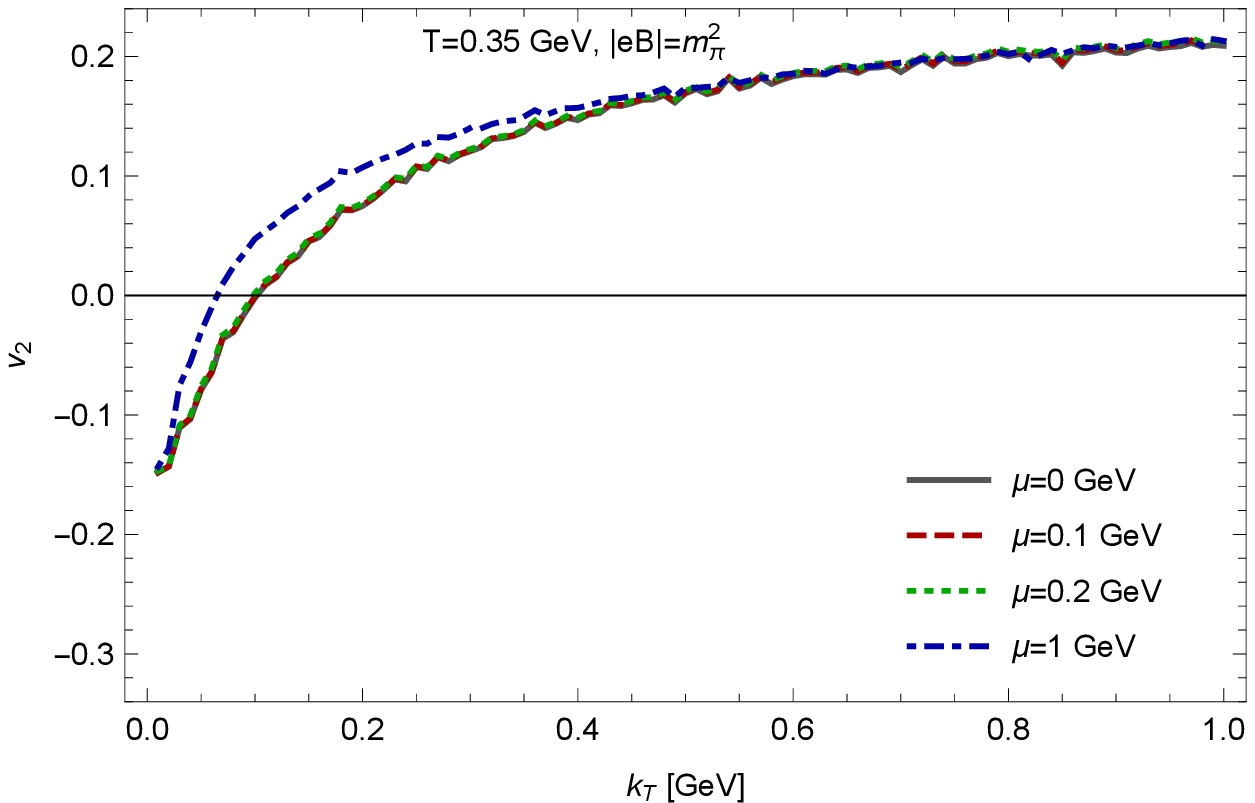}}
\subfigure[]{\includegraphics[width=0.45\textwidth]{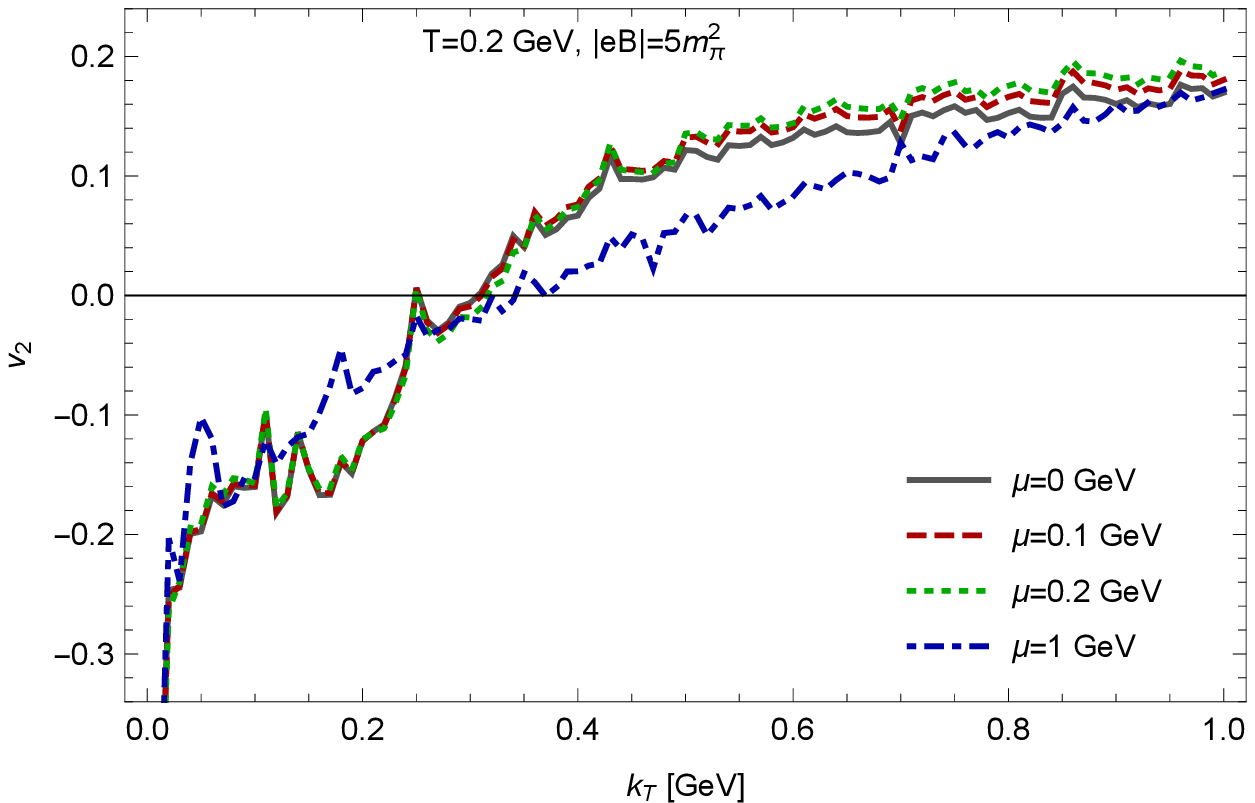}}
\hspace{0.01\textwidth}
\subfigure[]{\includegraphics[width=0.45\textwidth]{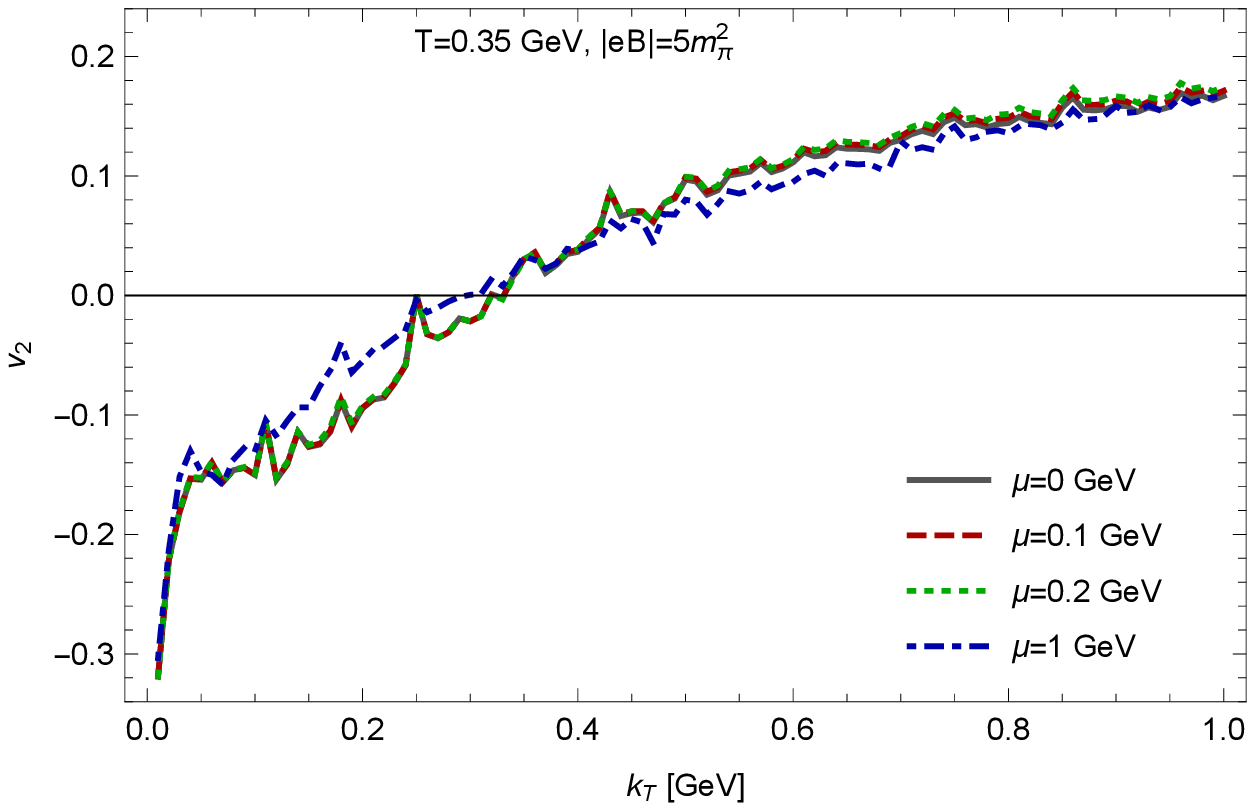}}
\caption{Ellipticity of photon emission as a function of the transverse momentum $k_T$ for four different values of the chemical potential: $\mu = 0$ (gray lines), $\mu = 0.1~\mbox{GeV}$ (red dashed lines), $\mu = 0.2~\mbox{GeV}$ (green dotted lines), and $\mu = 1~\mbox{GeV}$ (blue dot-dashed lines). The left panels (a and c) show the results for $T=200~\mbox{MeV}$ and the right panels (b and d) for $T=350~\mbox{MeV}$. The top and bottom panels show the results for $|eB| = m_{\pi}^2$ and $|eB| = 5 m_{\pi}^2$, respectively.}
\label{figure-v2}
\end{figure}

The ellipticity measure $v_2$ for the photon emission is shown as a function of the transverse momentum in Fig.~\ref{figure-v2}. As in the zero chemical potential case~\cite{Wang:2020dsr,Wang:2021ebh}, the value of $v_2$ is negative at small transverse momenta ($k_T\lesssim \sqrt{|eB|}$) and positive at at large momenta ($k_T\gtrsim \sqrt{|eB|}$). The former implies a stronger photon emission in the direction out of the reaction plane, while the latter implies a stronger photon emission along the reaction plane. Overall, the ellipticity is not affected much by a nonzero chemical potential even in the case of a very large $\mu$ ($\mu=1~\mbox{GeV}$). As in the $\mu=0$ case, the value of $v_2$ is of the order of $0.2$ at large transverse momenta. The weak dependence of $v_2$ on the chemical potential is probably explained by the fact that the rate is dominated by the same quark (and, to a lesser degree, the antiquark) splitting processes $q\to q+\gamma$ in almost all regimes. While the chemical potential affects the kinematics of the corresponding processes, the anisotropy of the corresponding synchrotron-like emission with respect to the magnetic field direction is largely the same.

\section{Summary and Conclusions}
\label{sec:Summary}

In this paper, we generalized the derivation of the photon polarization tensor in a strongly magnetized relativistic plasma to the case of a nonzero chemical potential. We found that the polarization tensor is determined by the same four symmetric and two antisymmetric structures that were identified in the $\mu=0$ study \cite{Wang:2021ebh}. However, while the antisymmetric parts of the tensor vanish at $\mu=0$, they are nonzero at $\mu\neq0$. This is the consequence of the charge conjugation symmetry breaking by the chemical potential. 

While the formal expressions are obtained for both real and imaginary parts of the polarization tensor, it is the absorptive part that was the main focus of this study. The latter is determined by the quark and antiquark splitting processes, $q\rightarrow q+\gamma$ and $\bar{q}\rightarrow \bar{q}+\gamma$, respectively, as well as the quark-antiquark annihilation, $q+\bar{q}\rightarrow \gamma$. Because of a broken charge conjugation symmetry at $\mu\neq0$, the two splitting processes give nonequal contributions. As expected, the quark (antiquark) splitting dominates over the antiquark (quark) one when the value of the baryon chemical potential is positive (negative). Also, the relative difference between the rates tends to go away with decreasing the chemical potential and increasing the temperature. As for the total photon production rate, it tends to grow with increasing of both temperature and chemical potential. We found that the contribution of the quark-antiquark annihilation remains small for a wide range of temperatures and chemical potentials relevant for heavy-ion collisions. In general, it gets larger with increasing the temperature and decreasing the chemical potential.

The ellipticity of the photon emission from a strongly magnetized quark-gluon plasma is not affected dramatically by a nonzero chemical potential. In fact, quantitative effects become noticeable only when $\mu$ is very large ($\sim 1~\mbox{GeV}$). This can be understood by recalling that a nonzero ellipticity is driven largely by the quark and antiquark splitting processes. While the relative weight of the two processes changes with $\mu$, their kinematics is not affected much by the chemical potential. As in the $\mu=0$ case, the photon emission is characterized by a negative ellipticity coefficient $v_2$ at small transverse momenta, $k_T\lesssim \sqrt{|eB|}$, and a positive $v_2$ at large momenta, $k_T\gtrsim \sqrt{|eB|}$. In other words, the profile of emission is  approximately prolate at small $k_T$ and oblate at large $k_T$. Because of the Landau-level quantization and the associated threshold effects, the ellipticity coefficient $v_2$ is neither smooth nor strictly monotonic function of the transverse momenta (energy). It is expected, however, that the interaction effects at subleading order will partially smooth out the corresponding dependence \cite{Wang:2021ebh}. 

In application to heavy-ion collisions, the findings of this study reconfirm that the photon emission from a strongly magnetized quark-gluon plasma is highly anisotropic \cite{Yee:2013qma,Tuchin:2014pka,Zakharov:2016mmc,Wang:2020dsr,Wang:2021ebh}. Such anisotropy, rather than a hydrodynamic flow of matter could explain a large $v_2$ in direct photon production observed in experiment~\cite{Adare:2011zr,Adare:2015lcd,Acharya:2018bdy}. As we show, a nonzero baryon chemical potential does not modify dramatically the existing theoretical predictions for the ellipticity of photon emission. Since the total rate is affected, however, the overall recalibration of the existing models might be needed.

In this study, similarly to other theoretical studies of the photon emission in the presence of a magnetic field, the calculations are done at the zeroth order in the strong coupling constant $\alpha_s$. This is a limitation that needs to be overcome before the qualitative conclusions about the ellipticity of emission are fully accepted. To this end, a systematic study of the gluon-mediated processes \cite{Kapusta:1991qp,Baier:1991em,Aurenche:1998nw,Steffen:2001pv,Arnold:2001ba,Arnold:2001ms,Ghiglieri:2013gia} has to be performed for a magnetized plasma. The corresponding generalization is quite challenging from a technical viewpoint however. Not only the complicated structure of the quark propagator complicates the problem, but also the resummation of the so-called hard ``magnetic loops" might be necessary \cite{Miransky:2002rp}.

\acknowledgements
The authors thank Shibing Chu and Kesheng Xu for providing computational resources. The authors also thank Liangkai Wu for useful suggestions on the numerical calculations. The work of X.W. was supported by the start-up funding No. 4111190010 of Jiangsu University and NSFC under Grant No. 11735007. The work of I.A.S. was supported by the U.S. National Science Foundation under Grant No.~PHY-1713950. 

\appendix
\section{Matsubara sums}
\label{MatsubaraSums}
In this Appendix, we present several general results for the fermionic Matsubara sums needed in the calculation of the photon polarization function at a nonzero chemical potential, i.e., 
\begin{eqnarray}
T\sum_{k=-\infty}^{\infty} 
\frac{1}{\left[(\omega_k-i \mu)^2+a^2\right]\left[(\omega_k-\Omega_m-i\mu)^2+b^2\right]} 
&=&\frac{\left[1-n_F(a-\mu)-n_F(b+\mu)\right]}{4ab\left(a+b - i \Omega_m\right)}+
\frac{\left[1-n_F(a+\mu)-n_F(b-\mu)\right]}{4ab\left(a+b + i \Omega_m\right)}\nonumber\\
&+&\frac{\left[n_F(a-\mu)-n_F(b-\mu)\right]}{4ab\left(a-b - i \Omega_m\right)}
+\frac{\left[n_F(a+\mu)-n_F(b+\mu)\right]}{4ab\left(a-b + i \Omega_m\right)},
\label{Matsubara-sum-1} \\
T\sum_{k=-\infty}^{\infty} 
\frac{(\omega_k-i\mu)(\omega_k-\Omega_m-i \mu)}{\left[(\omega_k-i \mu)^2+a^2\right]\left[(\omega_k-\Omega_m - i \mu)^2+b^2\right]} 
&=&\frac{\left[1-n_F(a-\mu)-n_F(b+\mu)\right]}{4\left(a+b - i \Omega_m\right)}+
\frac{\left[1-n_F(a+\mu)-n_F(b-\mu)\right]}{4\left(a+b + i \Omega_m\right)}\nonumber\\
&-&\frac{\left[n_F(a-\mu)-n_F(b-\mu)\right]}{4\left(a-b - i \Omega_m\right)}
-\frac{\left[n_F(a+\mu)-n_F(b+\mu)\right]}{4\left(a-b + i \Omega_m\right)},
\label{Matsubara-sum-2} \\
T\sum_{k=-\infty}^{\infty} \frac{i\omega_k + \mu}{\left[(\omega_k-i\mu)^2+a^2\right]\left[(\omega_k-\Omega_m-i\mu)^2+b^2\right]} 
&=&\frac{\left[1-n_F(a-\mu)-n_F(b+\mu)\right]}{4b\left(a+b - i \Omega_m\right)}-
\frac{\left[1-n_F(a+\mu)-n_F(b-\mu)\right]}{4b\left(a+b + i \Omega_m\right)}\nonumber\\
&+&\frac{\left[n_F(a-\mu)-n_F(b-\mu)\right]}{4b\left(a-b - i \Omega_m\right)}
-\frac{\left[n_F(a+\mu)-n_F(b+\mu)\right]}{4b\left(a-b + i \Omega_m\right)},
\label{Matsubara-sum-3} \\
T\sum_{k=-\infty}^{\infty} \frac{i(\omega_k-\Omega_m)+\mu}{\left[(\omega_k-i\mu)^2+a^2\right]\left[(\omega_k-\Omega_m-i\mu)^2+b^2\right]} 
&=&-\frac{\left[1-n_F(a-\mu)-n_F(b+\mu)\right]}{4a\left(a+b - i \Omega_m\right)}+
\frac{\left[1-n_F(a+\mu)-n_F(b-\mu)\right]}{4a\left(a+b + i \Omega_m\right)}\nonumber\\
&+&\frac{\left[n_F(a-\mu)-n_F(b-\mu)\right]}{4a\left(a-b - i \Omega_m\right)}
-\frac{\left[n_F(a+\mu)-n_F(b+\mu)\right]}{4a\left(a-b + i \Omega_m\right)},
\label{Matsubara-sum-4}
\end{eqnarray}
where $n_F(\epsilon) = 1/\left[\exp(\epsilon/T)+1\right]$ is the Fermi-Dirac distribution function, and  $\omega_k =  (2k+1)\pi T$ and $\Omega_m = 2m\pi T$ are the fermionic and bosonic Matsubara frequencies, respectively. Note that the distribution function satisfies the following relation: $n_F(-x)=1-n_F(x)$.

By making use of Eqs.~(\ref{Matsubara-sum-1}) through (\ref{Matsubara-sum-4}), it is straightforward to derive the following Matsubara sum of a more general type:  
\begin{eqnarray}
&& T\sum_{k=-\infty}^{\infty} 
\frac{(i\omega_k+\mu) (i\omega_k-i\Omega_m+\mu) X + (i\omega_k+\mu) Y_1 +(i\omega_k-i\Omega_m+\mu) Y_2 +Z}
{\left[(i\omega_k+\mu)^2-a^2\right]\left[(i\omega_k-i\Omega_m+\mu)^2-b^2\right]} \nonumber\\
&&= \sum_{\lambda=\pm 1} \sum_{\eta=\pm 1}
\frac{\left[ n_F(a+\eta \mu)-n_F(\lambda b+\eta \mu)  \right]}{4 \lambda ab\left(a-\lambda b+ \eta  i\Omega_m\right)}
\left[\lambda ab X-\eta\left( a Y_1+\lambda bY_2 \right)+Z\right],
\label{Matsubara-sum}
\end{eqnarray}
where coefficient $X$, $Y_1$, $Y_2$, and $Z$ are arbitrary functions of momenta. After replacing $a$ and $b$ with the Landau level energies, $E_{n,p_z,f}=\sqrt{p_z^2+m^2+2n|e_f B|}$ and $E_{n^{\prime},p_z-k_z,f}=\sqrt{(p_z-k_z)^2+m^2+2n^{\prime}|e_fB|}$, we obtain the Matsubara sums for the polarization function in the main text. 

\section{Explicit expressions for $I_{i,f}^{\mu\nu}$}
\label{AP-tr}

The four types of traces needed in the calculation of the polarization tensor are 
\begin{eqnarray}
\label{trt1}
T_{1,f}^{\mu\nu} &=&
\mbox{tr} \left[ \gamma^\mu \left(\mathcal{Q}_{\parallel} \gamma_\parallel + m \right)
 \left( {\cal P}_{+}L_n +{\cal P}_{-}L_{n-1} \right)
\gamma^\nu \left((\mathcal{Q}_{\parallel} -k_\parallel )\gamma_\parallel + m \right)
\left( {\cal P}_{+}L_{n^\prime} +{\cal P}_{-}L_{n^\prime-1} \right)\right]  
,\\
\label{trt2}
T_{2,f}^{\mu\nu} &=& \frac{i}{\ell_{f}^2}
\mbox{tr} \left[ \gamma^\mu \left(\mathcal{Q}_{\parallel} \gamma_\parallel + m \right)\left( {\cal P}_{+}L_n +{\cal P}_{-}L_{n-1} \right)
\gamma^\nu  (\mathbf{r}_{\perp}\cdot\bm{\gamma}_{\perp}) L_{n^\prime-1}^{1} \right] 
,\\
\label{trt3}
T_{3,f}^{\mu\nu} &=& -\frac{i}{\ell_{f}^2}
\mbox{tr} \left[ \gamma^\mu (\mathbf{r}_{\perp}\cdot\bm{\gamma}_{\perp}) L_{n-1}^{1}     
\gamma^\nu  \left((\mathcal{Q}_{\parallel} -k_\parallel )\gamma_\parallel + m \right)
\left( {\cal P}_{+}L_{n^\prime} +{\cal P}_{-}L_{n^\prime-1} \right)\right] 
,\\
\label{trt4}
T_{4,f}^{\mu\nu} &=&\frac{1}{\ell_{f}^4}
\mbox{tr} \left[ \gamma^\mu (\mathbf{r}_{\perp}\cdot\bm{\gamma}_{\perp}) L_{n-1}^{1}    
\gamma^\nu  (\mathbf{r}_{\perp}\cdot\bm{\gamma}_{\perp}) L_{n^\prime-1}^{1} \right],
\end{eqnarray}
where $\mathcal{Q}_\parallel \gamma_\parallel \equiv (p_0+\mu) \gamma^0 -p^{3}\gamma^3 = (i\omega_k+\mu)\gamma^0 -p^{3}\gamma^3$
and $(\mathcal{Q}_\parallel -k_\parallel )\gamma_\parallel \equiv (p_0+ \mu -k_0) \gamma^0 -(p^{3}-k^{3})\gamma^3 
= (i\omega_k+\mu-i\Omega_m)\gamma^0 -(p^{3}-k^{3})\gamma^3$.

After calculating the corresponding integrals over the transverse spatial coordinates, $\mathbf{r}_\perp$, the results are given by 
\begin{eqnarray}
I_{1,f}^{\mu\nu} &=& \int   d^2 \mathbf{r}_\perp e^{-i \mathbf{r}_\perp\cdot \mathbf{k}_\perp} e^{-\mathbf{r}_\perp^2/(2\ell_{f}^2)} T_{1,f}^{\mu\nu} =-4 \pi \ell_{f}^2 g_\perp^{\mu\nu} \left[(\mathcal{Q}_\parallel  ((\mathcal{Q}_\parallel -k_\parallel )-m^2 \right]
\left[ \mathcal{I}_{0,f}^{n,n^{\prime}-1}(\xi)+\mathcal{I}_{0,f}^{n-1,n^{\prime}}(\xi) \right]
 \nonumber\\
&+&4 \pi \ell_{f}^2 \left[\mathcal{Q}_\parallel^\mu (\mathcal{Q}_\parallel -k_\parallel )^\nu+(\mathcal{Q}_\parallel -k_\parallel )^\mu \mathcal{Q}_\parallel^\nu 
-g_\parallel^{\mu\nu} \left[\mathcal{Q}_\parallel  (\mathcal{Q}_\parallel -k_\parallel )-m^2 \right] \right]
\left[\mathcal{I}_{0,f}^{n,n^{\prime}}(\xi)+\mathcal{I}_{0,f}^{n-1,n^{\prime}-1}(\xi) \right] \nonumber\\
&+& 4 \pi i \ell_{f}^4 e_f F^{\mu\nu}
\left[\mathcal{Q}_\parallel  (\mathcal{Q}_\parallel -k_\parallel )-m^2 \right]
\left[\mathcal{I}_{0,f}^{n,n^{\prime}-1}(\xi)-\mathcal{I}_{0,f}^{n-1,n^{\prime}}(\xi)\right] , 
\label{I_if-1-appD}
\\
I_{2,f}^{\mu\nu}&=& \int   d^2 \mathbf{r}_\perp e^{-i \mathbf{r}_\perp\cdot \mathbf{k}_\perp} e^{-\mathbf{r}_\perp^2/(2\ell_{f}^2)} T_{2,f}^{\mu\nu}  
= - 4\pi \ell_{f} \left(\mathcal{Q}_\parallel^\mu \hat{k}_\perp^{\nu}+ \hat{k}_\perp^{\mu}\mathcal{Q}_\parallel^\nu \right)
\left[\mathcal{I}_{1,f}^{n,n^{\prime}-1}(\xi)+\mathcal{I}_{1,f}^{n-1,n^{\prime}-1}(\xi)\right] 
\nonumber\\
&+& 4 \pi i \ell_{f}^3 e_f \left(\mathcal{Q}_\parallel^\mu F^{\nu\rho} \hat{k}_{\perp,\rho}  -   F^{\mu\rho}\hat{k}_{\perp,\rho}\mathcal{Q}_\parallel^\nu \right)
\left[\mathcal{I}_{1,f}^{n,n^{\prime}-1}(\xi)-\mathcal{I}_{1,f}^{n-1,n^{\prime}-1}(\xi) \right],
\label{I_if-2-appD}
\\
I_{3,f}^{\mu\nu}&=& \int d^2 \mathbf{r}_\perp e^{-i \mathbf{r}_\perp\cdot \mathbf{k}_\perp} e^{-\mathbf{r}_\perp^2/(2\ell_{f}^2)} T_{3,f}^{\mu\nu} = 4\pi \ell_{f} 
\left((\mathcal{Q}_\parallel -k_\parallel )^{\mu} \hat{k}_\perp^{\nu}+ \hat{k}_\perp^{\mu}(\mathcal{Q}_\parallel -k_\parallel )^\nu \right)
\left[ \mathcal{I}_{1,f}^{n^{\prime},n-1}(\xi)+\mathcal{I}_{1,f}^{n^{\prime}-1,n-1}(\xi)\right]
\nonumber\\
&+& 4\pi i \ell_{f}^3 e_f
\left( (\mathcal{Q}_\parallel -k_\parallel )^\mu F^{\nu\rho} \hat{k}_{\perp,\rho} - F^{\mu\rho}\hat{k}_{\perp,\rho}(\mathcal{Q}_\parallel -k_\parallel )^\nu \right)
\left[ \mathcal{I}_{1,f}^{n^{\prime},n-1}(\xi)-\mathcal{I}_{1,f}^{n^{\prime}-1,n-1}(\xi) \right], 
\label{I_if-3-appD}
\\
I_{4,f}^{\mu\nu} &=& \int   d^2 \mathbf{r}_\perp e^{-i \mathbf{r}_\perp\cdot \mathbf{k}_\perp} e^{-\mathbf{r}_\perp^2/(2\ell_{f}^2)} T_{4,f}^{\mu\nu}
= 8\pi \left[ g_{\parallel}^{\mu\nu}  \mathcal{I}_{2,f}^{n-1,n^{\prime}-1}(\xi)
-\left(g_{\perp}^{\mu\nu}+ 2\hat{k}_\perp^{\mu} \hat{k}_\perp^{\nu}  \right)
\mathcal{I}_{3,f}^{n-1,n^{\prime}-1}(\xi)
 \right],
\label{I_if-4-appD}
\end{eqnarray}
where $\xi =k_\perp^2 \ell_{f}^2/2$.
The explicit expressions for functions $\mathcal{I}_{i,f}^{n,n^{\prime}}(\xi)$ ($i = 1, 2, 3, 4$) are given in~\cite{Wang:2021ebh} 
\begin{eqnarray}
\mathcal{I}_{0,f}^{n,n^{\prime}}(\xi)&=& (-1)^{n+n^\prime} e^{-\xi} L_{n}^{n^\prime-n}\left(\xi\right) 
L_{n^\prime}^{n-n^\prime}\left(\xi\right) ,
\label{I0f-LL}  \\
\mathcal{I}_{1,f}^{n,n^{\prime}}(\xi)&=& \sqrt{2\xi} (-1)^{n+n^\prime} e^{-\xi} L_{n}^{n^\prime-n+1}\left(\xi\right) 
L_{n^\prime}^{n-n^\prime}\left(\xi\right) 
\label{I1f-LL} , \\
\mathcal{I}_{2,f}^{n,n^{\prime}}(\xi)&=&2 (-1)^{n+n^\prime}(n^\prime+1)
e^{-\xi} L_{n}^{n^\prime-n}\left(\xi\right) 
L_{n^\prime+1}^{n-n^\prime}\left(\xi\right) , 
\label{I2f-LL} \\
\mathcal{I}_{3,f}^{n,n^{\prime}}(\xi)&=&  2(-1)^{n+n^\prime} \xi
e^{-\xi} L_{n}^{n^\prime-n+1}\left(\xi\right) 
L_{n^\prime}^{n-n^\prime+1}\left(\xi\right).
\label{I3f-LL} 
\end{eqnarray}
For properties and relations that these functions satisfy, see Ref.~\cite{Wang:2021ebh}.

After performing the Matsubara sums, making the analytical continuation $i\Omega_m\to \Omega+i \epsilon$, and using the energy conservation condition ($E_{n,p_z,f}-\lambda E_{n^{\prime},p_z-k_z,f}+ \eta\Omega=0$), one finds that the $I_{i,f}^{\mu\nu}(\xi)$ functions in the final result will be replaced by analogous expressions, where only the following replacements are made:
\begin{eqnarray}
\mathcal{Q}_\parallel ^\mu & \to &  \bar{\mathcal{Q}}_\parallel^\mu  \Big|_{p_z=p_{z,f}^{(\pm)}}
 = -\eta E_{n,p_z,f} \delta^{\mu}_{0} +p_z\delta^{\mu}_{3} \Big|_{p_z=p_{z,f}^{(\pm)}}
 \nonumber\\
 &=&\frac{1}{2}k_\parallel^\mu\left(\frac{2(n-n^{\prime})|e_f B|}{\Omega^2-k_z^2} +1\right)
 \pm\frac{1}{2}\tilde{k}_\parallel^\mu\sqrt{ \left(1-\frac{(k_{-}^f)^2}{\Omega^2-k_z^2} \right)
\left( 1-\frac{(k_{+}^f)^2}{\Omega^2-k_z^2}\right)}
\label{Q-after-sum}  , \\
 (\mathcal{Q}_\parallel -k_\parallel )^\mu & \to &  (\bar{\mathcal{Q}}_\parallel -k_\parallel )^\mu   \Big|_{p_z=p_{z,f}^{(\pm)}}
 = -\eta \lambda E_{n^{\prime},p_z-k_z,f} \delta^{\mu}_{0} +(p_z-k_z)\delta^{\mu}_{3}   \Big|_{p_z=p_{z,f}^{(\pm)}}
  \nonumber\\
 &=&\frac{1}{2}k_\parallel^\mu\left(\frac{2(n-n^{\prime})|e_f B|}{\Omega^2-k_z^2} -1\right)
  \pm\frac{1}{2}\tilde{k}_\parallel^\mu\sqrt{ \left(1-\frac{(k_{-}^f)^2}{\Omega^2-k_z^2} \right)
\left( 1-\frac{(k_{+}^f)^2}{\Omega^2-k_z^2}\right)}
\label{Q-k-after-sum}  ,\\
 \mathcal{Q}_\parallel  (\mathcal{Q}_\parallel -k_\parallel) & \to &   \bar{\mathcal{Q}}_\parallel (\bar{\mathcal{Q}}_\parallel -k_\parallel ) \Big|_{p_z=p_{z,f}^{(\pm)}}
 =  \lambda E_{n,p_z,f}E_{n^{\prime},p_z-k_z,f} -p_z(p_z-k_z) \Big|_{p_z=p_{z,f}^{(\pm)}} 
  \nonumber\\
 &=&  m^2+(n+n^\prime)|e_f B| -\frac{1}{2} k_\parallel^2 ,
 \label{Q-Q-k-after-sum} 
\end{eqnarray}
where $k_\parallel^\mu  = \Omega \delta^{\mu}_{0} +k_z \delta^{\mu}_{3} $ and 
$\tilde{k}_\parallel^\mu  = k_z \delta^{\mu}_{0} +\Omega \delta^{\mu}_{3} $.
Note that $k_{\parallel,\mu} \tilde{k}_\parallel^\mu =0$ and $\tilde{k}_{\parallel,\mu} \tilde{k}_\parallel^\mu = - k_\parallel^2$. The definition of the transverse threshold momenta $k_{\pm}$ are given in Eq.~(\ref{kpm}).
Note that there is no dependence on the chemical potential in Eqs.~(\ref{Q-after-sum}) -- (\ref{Q-Q-k-after-sum}). It is the consequence of the general result for the Matsubara sum in Eq.~(\ref{Matsubara-sum}).

By using the above results, we can derive the expressions for $I_{i,f}^{\mu\nu}$ tensors, which have the same form as in the $\mu=0$ case~\cite{Wang:2021ebh}, i.e.,
\begin{subequations}
\label{I_if-appD}
 \begin{eqnarray}
I_{1,f}^{\mu\nu}\Big|_{p_z=p_{z,f}^{(\pm)}}  &=& -4 \pi \ell_{f}^2 g_\perp^{\mu\nu} \left[ (n+n^\prime)|e_{f} B| -\frac{1}{2} k_\parallel^2 \right]
\left[ \mathcal{I}_{0,f}^{n,n^{\prime}-1}(\xi)+\mathcal{I}_{0,f}^{n-1,n^{\prime}}(\xi) \right]
\nonumber\\
&- &4 \pi \ell_{f}^2 g_\parallel^{\mu\nu} \left[ (n+n^\prime)|e_{f} B| -\frac{1}{2} k_\parallel^2 \right] 
\left[\mathcal{I}_{0,f}^{n,n^{\prime}}(\xi)+\mathcal{I}_{0,f}^{n-1,n^{\prime}-1}(\xi) \right]
\nonumber\\
&+& 4 \pi i \ell_{f}^4  e_f F^{\mu\nu}
\left[ (n+n^\prime)|e_{f} B| -\frac{1}{2} k_\parallel^2 \right]
\left[\mathcal{I}_{0,f}^{n,n^{\prime}-1}(\xi)-\mathcal{I}_{0,f}^{n-1,n^{\prime}}(\xi)\right] 
\nonumber\\
&+& 4 \pi \ell_{f}^2 A_{\pm}^{\mu \nu}
\left[\mathcal{I}_{0,f}^{n,n^{\prime}}(\xi)+\mathcal{I}_{0,f}^{n-1,n^{\prime}-1}(\xi) \right],
\label{I_if-1-appD}
\end{eqnarray}
\begin{eqnarray}
I_{2,f}^{\mu\nu}+ I_{3,f}^{\mu\nu}\Big|_{p_z=p_{z,f}^{(\pm)}}  &=& - 4\pi \ell_{f} B_{\pm}^{\mu \nu}
\left[\mathcal{I}_{1,f}^{n,n^{\prime}-1}(\xi)+\mathcal{I}_{1,f}^{n-1,n^{\prime}-1}(\xi)\right] 
\nonumber\\
&-& 4\pi \ell_{f} 
\left(k_{\parallel}^{\mu}\hat{k}_{\perp}^{\nu}+k_{\parallel}^{\nu}\hat{k}_{\perp}^{\mu} -B_{\pm}^{\mu \nu} \right)
\left[ \mathcal{I}_{1,f}^{n^{\prime},n-1}(\xi)+\mathcal{I}_{1,f}^{n^{\prime}-1,n-1}(\xi)\right] 
\nonumber\\
&-& 4 \pi i \ell_{f}^3 \frac{e_{f}B}{k_\perp} C_{\pm}^{\mu \nu}
\left[\mathcal{I}_{1,f}^{n,n^{\prime}-1}(\xi)-\mathcal{I}_{1,f}^{n-1,n^{\prime}-1}(\xi) \right]
\nonumber\\
&-& 4\pi i \ell_{f}^3  \frac{e_{f}B}{k_\perp} 
\left[ C_{\pm}^{\mu \nu} - \left( k_{\parallel}^{\mu}\tilde{k}_{\perp}^{\nu}-k_{\parallel}^{\nu}\tilde{k}_{\perp}^{\mu}\right) \right]
\left[ \mathcal{I}_{1,f}^{n^{\prime},n-1}(\xi)-\mathcal{I}_{1,f}^{n^{\prime}-1,n-1}(\xi) \right],
\label{I_if-2+3-appD}
\end{eqnarray}
\begin{eqnarray}
I_{4,f}^{\mu\nu}\Big|_{p_z=p_{z,f}^{(\pm)}}  &=&  8\pi \left[ g_{\parallel}^{\mu\nu}  \mathcal{I}_{2,f}^{n-1,n^{\prime}-1}(\xi)
-\left(g_{\perp}^{\mu\nu}+ 2\hat{k}_\perp^{\mu} \hat{k}_\perp^{\nu}  \right)
\mathcal{I}_{3,f}^{n-1,n^{\prime}-1}(\xi)
 \right].
 \label{I_if-4-appD}
\end{eqnarray}
\end{subequations}
Note that the upper and lower signs correspond to $p_{z,f}^{(+)}$ and $p_{z,f}^{(-)}$, respectively. Also we used the following shorthand notation:
\begin{subequations}
\begin{eqnarray}
A_{\pm}^{\mu \nu}
&=& -\frac{1}{2}k_\parallel^\mu k_\parallel^\nu
\left( 1- \frac{4(n-n^\prime)^2 (e_{f}B)^2}{(\Omega^2-k_z^2)^2} \right)
+\frac{1}{2} \tilde{k}_{\parallel}^{\mu}\tilde{k}_{\parallel}^{\nu}
\left( 1-\frac{4\left[m^2+(n+n^\prime)|e_{f} B|\right]}{\Omega^2-k_z^2} + \frac{4(n-n^\prime)^2 (e_{f}B)^2}{(\Omega^2-k_z^2)^2} \right)
\nonumber\\
&\pm & \left( k_{\parallel}^{\mu}\tilde{k}_{\parallel}^{\nu}+k_{\parallel}^{\nu}\tilde{k}_{\parallel}^{\mu}\right)
\frac{(n-n^{\prime})|e_{f} B|}{\Omega^2-k_z^2} 
\sqrt{1-\frac{4\left[m^2+(n+n^\prime)|e_{f} B|\right]}{\Omega^2-k_z^2} + \frac{4(n-n^\prime)^2 (e_{f}B)^2}{(\Omega^2-k_z^2)^2}} ,
\end{eqnarray}
\begin{equation}
B_{\pm}^{\mu \nu}
= \frac{1}{2}\left( k_{\parallel}^{\mu}\hat{k}_{\perp}^{\nu}+k_{\parallel}^{\nu}\hat{k}_{\perp}^{\mu}\right)
\left(\frac{2(n-n^{\prime})|e_{f} B|}{\Omega^2-k_z^2} + 1 \right)
\pm \frac{1}{2}\left( \tilde{k}_{\parallel}^{\mu}\hat{k}_{\perp}^{\nu}+\tilde{k}_{\parallel}^{\nu}\hat{k}_{\perp}^{\mu}\right)
\sqrt{1-\frac{4\left[m^2+(n+n^\prime)|e_{f} B|\right]}{\Omega^2-k_z^2} + \frac{4(n-n^\prime)^2 (e_{f}B)^2}{(\Omega^2-k_z^2)^2}} ,
\end{equation}
\begin{equation}
C_{\pm}^{\mu \nu}
= \frac{1}{2}\left( k_{\parallel}^{\mu}\tilde{k}_{\perp}^{\nu}-k_{\parallel}^{\nu}\tilde{k}_{\perp}^{\mu}\right)
\left(\frac{2(n-n^{\prime})|e_{f} B|}{\Omega^2-k_z^2} + 1 \right)
\pm \frac{1}{2}\left( \tilde{k}_{\parallel}^{\mu}\tilde{k}_{\perp}^{\nu}-\tilde{k}_{\parallel}^{\nu}\tilde{k}_{\perp}^{\mu}\right)
\sqrt{1-\frac{4\left[m^2+(n+n^\prime)|e_{f} B|\right]}{\Omega^2-k_z^2} + \frac{4(n-n^\prime)^2 (e_{f}B)^2}{(\Omega^2-k_z^2)^2}}.
\end{equation}
\end{subequations}
Now, let us introduce the Lorentz contracted expression functions: $\mathcal{F}_i^{f} = g_{\mu\nu}I_{i,f}^{\mu\nu}$,  by using the definitions in Eq.~(\ref{I_if-appD}), one finds that
\begin{eqnarray}
\mathcal{F}_1^{f}&=&  8\pi \left[\frac{k_\parallel^2\ell_{f}^2}{2} -(n+n^\prime)\right]
\left( \mathcal{I}_{0,f}^{n-1,n^{\prime}}(\xi) +\mathcal{I}_{0,f}^{n,n^{\prime}-1}(\xi)  \right)
+8\pi \ell_{f}^2 m^2\left( \mathcal{I}_{0,f}^{n,n^{\prime}}(\xi) +\mathcal{I}_{0,f}^{n-1,n^{\prime}-1}(\xi)  \right)
\label{F1-orig} ,\\
\mathcal{F}_2^{f}&=& g_{\mu \nu} I_{2,f}^{\mu \nu}=0, \\
\mathcal{F}_3^{f}&=& g_{\mu \nu} I_{3,f}^{\mu \nu}=0,  \\
\mathcal{F}_4^{f}&=& 16 \pi \, \mathcal{I}_{2,f}^{n-1,n^{\prime}-1}(\xi).
\label{F4-orig}
\end{eqnarray}
By adding together the nonvanishing functions $\mathcal{F}_i^{f}$, we obtain 
\begin{equation}
\mathcal{F}_1^{f} + \mathcal{F}_4^{f} 
= 8\pi \left(n+n^{\prime}+m^2\ell_{f}^2\right)\left[\mathcal{I}_{0,f}^{n,n^{\prime}}(\xi)+\mathcal{I}_{0,f}^{n-1,n^{\prime}-1}(\xi) \right]
+8\pi  \left(\frac{k_{\parallel}^2 -k_{\perp}^2}{2} \ell_{f}^2  -(n+n^{\prime})\right)
\left[\mathcal{I}_{0,f}^{n,n^{\prime}-1}(\xi)+\mathcal{I}_{0,f}^{n-1,n^{\prime}}(\xi) \right]. 
\end{equation}
Note that the second term simplifies when photons satisfy the on-shell condition $k_{\parallel}^2 = k_{\perp}^2 $.

\section{Tensor structure of $\mbox{Im} \left[\Pi^{\mu\nu} \right] $}
\label{app:Tensor-structure}

By making use of the expression for the polarization tensor in Eq.~(\ref{Im-Pol-fun}) and the definition of tensors $I_{i,f}^{\mu\nu}$ in Appendix~\ref{AP-tr}, we find that the imaginary part of $\Pi_R^{\mu\nu}(\Omega;\mathbf{k})$ has the following structure:
\begin{eqnarray}
\mbox{Im} \left[\Pi_R^{\mu\nu}(\Omega;\mathbf{k}) \right] &=&
  \left( \frac{k_\parallel^\mu k_\parallel^\nu }{k_\parallel^2} -  g_{\parallel}^{\mu\nu} \right)  \mbox{Im} \left[\Pi_{1}\right]
+ \left(g_{\perp}^{\mu\nu}+ \frac{k_\perp^\mu k_\perp^\nu}{k_\perp^2} \right) \mbox{Im} \left[\Pi_{2}\right]
\nonumber\\
&+& \left( \frac{k_\parallel^\mu \tilde{k}_\parallel^\nu +\tilde{k}_\parallel^\mu k_\parallel^\nu}{k_\parallel^2}  
+ \frac{\tilde{k}_\parallel^\mu k_\perp^\nu +k_\perp^\mu \tilde{k}_\parallel^\nu}{k_\perp^2}\right) \mbox{Im} \left[\Pi_{3}\right]
+\left(\frac{k_\parallel^\mu k_\perp^\nu +k_\perp^\mu k_\parallel^\nu}{k_\parallel^2 } + \frac{k_\perp^2}{k_\parallel^2 }  g_{\parallel}^{\mu\nu}  - g_{\perp}^{\mu\nu} \right) \mbox{Im} \left[\Pi_{4} \right] \nonumber\\
&+& \left( \frac{F^{\mu\nu} }{B}
+ \frac{k_\parallel^\mu \tilde{k}_{\perp}^\nu -\tilde{k}_{\perp}^\mu k_\parallel^\nu}{k_\parallel^2} \right) \mbox{Im} \left[\tilde{\Pi}_{5}\right]
+ \frac{\tilde{k}_\parallel^\mu \tilde{k}_{\perp}^\nu -\tilde{k}_{\perp}^\mu \tilde{k}_\parallel^\nu}{k_\parallel^2} \mbox{Im} \left[\tilde{\Pi}_{6}\right] ,
\label{ImPi-tensor-app0}
\end{eqnarray}
where we utilized the shorthand notations introduced in Eq.~(\ref{ggkkkk}). Note that $k_\mu \tilde{k}_{\perp}^{\mu} =k_{\mu} \tilde{k}_\parallel^\mu =0$ and $F^{\mu\nu} k_{\perp,\nu} =B \tilde{k}_{\perp}^{\mu}$. One can also check that 
$\tilde{k}_{\perp,\mu} \tilde{k}_{\perp}^{\mu} = k_{\perp,\mu} k_{\perp}^{\mu} = -k_\perp^2$ and 
$\tilde{k}_{\parallel,\mu} \tilde{k}_\parallel^\mu = -k_{\parallel,\mu} k_\parallel^\mu= - k_\parallel^2$.

To simplify the representation of the six component functions in the imaginary part of polarization tensor (\ref{ImPi-tensor-app0}), it is convenient to introduce the following operator:
\begin{equation}
\hat{\cal X} (\ldots)=
 \sum_{f=u, d} \frac{N_c \alpha_f }{4\pi \ell_{f}^4} \sum_{n,n^\prime=0}^{\infty} 
\sum_{\lambda,\eta=\pm 1}\sum_{s=\pm 1} \Theta_{\lambda, \eta}^{n,n^{\prime}}(\Omega,k_z)
\frac{n_F(E_{n,p_z,f}+\eta \mu)-n_F(\lambda E_{n^{\prime},p_z-k_z,f}+\eta \mu) }{\eta\lambda  \sqrt{ \left( \Omega^2-k_z^2-(k_{-}^f)^2 \right) \left( \Omega^2-k_z^2-(k_{+}^f)^2\right)} } (\ldots).
\label{hat-X-app}
\end{equation} 
Note that, with the help of the operator $\hat{\cal X}$, the polarization tensor in Eq.~(\ref{Im-Pol-fun}) can be rewritten in a compact form as $\mbox{Im} \left[\Pi_R^{\mu\nu}\right]  = \hat{\cal X}  \sum_{i=1}^{4}I_{i,f}^{\mu\nu} $. 

By using the operator in Eq.~(\ref{hat-X-app}), we can also write down the explicit expressions for the individual tensor component functions. In particular, the first four components, defining the symmetric tensor structures, are
\begin{eqnarray}
\mbox{Im} \left[\Pi_{1}\right] &=& 
8 \pi \hat{\cal X} \left(\frac{2(n-n^{\prime})^2}{k_\parallel^2\ell_{f}^2} -m^2\ell_{f}^2-(n+n^\prime)\right)
\left[\mathcal{I}_{0,f}^{n,n^{\prime}}(\xi)+\mathcal{I}_{0,f}^{n-1,n^{\prime}-1}(\xi) \right], 
\\
\mbox{Im} \left[\Pi_{2}\right] &=& -16 \pi \hat{\cal X} \, \mathcal{I}_{3,f}^{n-1,n^{\prime}-1}(\xi)
\nonumber\\
&=& 
-8 \pi \hat{\cal X}  \Bigg\{
(n+n^{\prime})\left[\mathcal{I}_{0,f}^{n,n^{\prime}-1}(\xi) +\mathcal{I}_{0,f}^{n-1,n^{\prime}}(\xi) \right]
-\frac{(n-n^{\prime})^2}{\xi}\left[\mathcal{I}_{0,f}^{n,n^{\prime}}(\xi) +\mathcal{I}_{0,f}^{n-1,n^{\prime}-1}(\xi) \right] 
\Bigg\}, 
\\
\mbox{Im} \left[\Pi_{3}\right] &=& \pm 4 \pi \hat{\cal X}  (n-n^{\prime}) 
\sqrt{\left(1-\frac{k_{-}^2}{k_\parallel^2}\right)\left(1-\frac{k_{+}^2}{k_\parallel^2}\right)}
\left[\mathcal{I}_{0,f}^{n,n^{\prime}}(\xi)+\mathcal{I}_{0,f}^{n-1,n^{\prime}-1}(\xi) \right],
\\
\mbox{Im} \left[\Pi_{4}\right] &=& -2\pi \hat{\cal X} \Bigg\{ \frac{k_\parallel^2\ell_{f}^2+2(n-n^{\prime})}{k_\perp \ell_{f}}
\left[\mathcal{I}_{1,f}^{n,n^{\prime}-1}(\xi)+\mathcal{I}_{1,f}^{n-1,n^{\prime}-1}(\xi)\right] 
+  \frac{ k_\parallel^2\ell_{f}^2 - 2(n-n^{\prime})}{k_\perp \ell_{f}}
\left[ \mathcal{I}_{1,f}^{n^{\prime},n-1}(\xi)+\mathcal{I}_{1,f}^{n^{\prime}-1,n-1}(\xi)\right] \Bigg\}
\nonumber\\
&=& - 2\pi \hat{\cal X}  \left\{k_\parallel^2\ell_{f}^2
\left[\mathcal{I}_{0,f}^{n,n^{\prime}-1}(\xi)+\mathcal{I}_{0,f}^{n-1,n^{\prime}}(\xi) \right]
-\frac{2 (n-n^{\prime})^2 }{\xi}
\left[\mathcal{I}_{0,f}^{n,n^{\prime}}(\xi)+\mathcal{I}_{0,f}^{n-1,n^{\prime}-1}(\xi) \right] \right\}.
\end{eqnarray}
Similarly, the component functions of the antisymmetric contributions read 
\begin{eqnarray}
\mbox{Im} \left[\tilde{\Pi}_{5}\right]  &=& -  2 \pi i s_\perp \hat{\cal X} \left( k_\parallel^2 \ell_{f}^2 - 2(n+n^\prime) \right)
\left[\mathcal{I}_{0,f}^{n,n^{\prime}-1}(\xi)-\mathcal{I}_{0,f}^{n-1,n^{\prime}}(\xi)\right] ,
\end{eqnarray}
and 
\begin{eqnarray}
\mbox{Im} \left[\tilde{\Pi}_{6}\right]  &=&  \mp 2 \pi i \hat{\cal X}  \frac{s_\perp  k_\parallel^2 \ell_{f}^2}{k_\perp\ell_{f}}  \sqrt{ \left(1-\frac{k_{-}^2}{k_\parallel^2} \right)
\left( 1-\frac{k_{+}^2}{k_\parallel^2}\right)}
\left[\mathcal{I}_{1,f}^{n,n^{\prime}-1}(\xi)-\mathcal{I}_{1,f}^{n-1,n^{\prime}-1}(\xi) 
+ \mathcal{I}_{1,f}^{n^{\prime},n-1}(\xi)-\mathcal{I}_{1,f}^{n^{\prime}-1,n-1}(\xi) \right]
\nonumber\\
&=& \mp 4\pi i   \frac{s_{\perp}}{k_\perp^2}  \hat{\cal X}  (n+n^\prime) \sqrt{ \left(k_\parallel^2- k_{-}^2  \right)
\left(k_\parallel^2- k_{+}^2 \right)}
\left[\mathcal{I}_{0,f}^{n,n^{\prime}}(\xi)-\mathcal{I}_{0,f}^{n-1,n^{\prime}-1}(\xi)\right]. 
\end{eqnarray}

\end{document}